# Statistics and Topology of the *COBE*[1] DMR First Year Sky Maps

G. F. Smoot[2], L. Tenorio[2], A. J. Banday[3], A. Kogut[4], E. L. Wright[5], G. Hinshaw[4], and C. L. Bennett[6]

[1] The National Aeronautics and Space Administration/Goddard Space Flight Center (NASA/GSFC) is responsible for the design, development, and operation of the Cosmic Background Explorer (*COBE*). Scientific guidance is provided by the *COBE* Science Working Group. GSFC is also responsible for the development of the analysis software and for the production of the mission data sets.

[2] LBL, SSL, & CfPA, Bldg 50-351, University of California, Berkeley CA 94720

[3] University Space Research Assoc., Code 685.9 NASA/GSFC, Greenbelt MD 20771

[4] Hughes STX Corporation, Code 685.9 NASA/GSFC, Greenbelt MD 20771

[5] UCLA Astronomy Department, Los Angeles CA 90024-1562

[6] NASA Goddard Space Flight Center, Code 685, Greenbelt MD 20771

## ABSTRACT

We use statistical and topological quantities to test the *COBE*-DMR first year sky maps against the hypothesis that the observed temperature fluctuations reflect Gaussian initial density perturbations with random phases. Recent papers discuss specific quantities as discriminators between Gaussian and non-Gaussian behavior, but the treatment of instrumental noise on the data is largely ignored. The presence of noise in the data biases many statistical quantities in a manner dependent on both the noise properties and the unknown CMB temperature field. Appropriate weighting schemes can minimize this effect, but it cannot be completely eliminated. Analytic expressions are presented for these biases, and Monte Carlo simulations used to assess the best strategy for determining cosmologically interesting information from noisy data. The genus is a robust discriminator that can be used to estimate the power law quadrupole-normalized amplitude, $Q_{rms-PS}$, independently of the 2-point correlation function. The genus of the DMR data are consistent with Gaussian initial fluctuations with $Q_{rms-PS} = (15.7 \pm 2.2) - (6.6 \pm 0.3)(n - 1)$ $\mu$K where $n$ is the power law index. Fitting the rms temperature variations at various smoothing angles gives $Q_{rms-PS} = 13.2 \pm 2.5$ $\mu$K and $n = 1.7^{+0.3}_{-0.6}$. While consistent with Gaussian fluctuations, the first year data are only sufficient to rule out strongly non-Gaussian distributions of fluctuations.

# 1   Introduction

The origin of large scale structure in the Universe is one of the most fundamental issues in cosmology. Gravitational instability models hold that large-scale structure forms as the result of gravitational amplification of initially small perturbations in the primordial density distribution. These models commonly assume Gaussian initial fluctuations with amplitude described by a nearly scale-invariant primordial power spectrum. The inflationary model of the early Universe produces such a primordial power spectrum of density fluctuations (Guth & Pi 1982, Starobinskii 1982, Hawking 1982, Bardeen, Steinhardt & Turner 1983). Structure formation mechanisms that involve non-Gaussian fluctuations include topological defects such as global monopoles (Bennett & Rhie 1993), cosmic strings (Vilenkin 1985, Stebbins et al. 1987), domain walls (Stebbins & Turner 1989, Turner, Watkins, & Widrow 1991), and textures (Turok 1989, Turok & Spergel 1990). Late-time phase transitions (e.g. Hill, Schramm & Fry 1989) and axions (e.g. Kolb & Turner 1989) produce structure in the Universe through dynamical effects and also generate non-Gaussian density fluctuations.

   The determination of the nature of the initial density fluctuations is an important constraint to cosmological models. Unfortunately, non-trivial bias is expected in the process of galaxy formation as a consequence of the gravitational evolution of the initial Gaussian seeds in the non-linear regime (Fry & Gastañaga, 1993), so the observed statistics of the galaxy distribution do not necessarily reflect the underlying density fluctuation distribution. Whether the initial fluctuations were Gaussian or non-Gaussian might eventually be resolved by analysis of the cosmic microwave background (CMB), particularly on angular scales larger than a few degrees where the observations directly reflect the initial density perturbations.

   Several recent papers have considered the statistical analysis of CMB observations, and the limitations imposed by the cosmic variance (Graham et al., 1993; Xiaochun & Schramm, 1993a,b; Cayón et al., 1991; White, Krauss, & Silk 1993, Crittenden et al. 1993; Srednicki, 1993). Instrumental noise in the CMB data is also a problem, capable of generating biases in statistical quantities in a fashion dependent on the noise properties and the unknown CMB temperature field. Understanding these effects is crucial to the interpretation of the statistics of CMB maps. We present analytic expressions for these biases and use Monte Carlo simulations to determine the most useful strategy for recovering cosmologically interesting quantities from the data.

   Gott et al. (1990) suggested that the genus is a useful discriminator for non-Gaussian statistics. We present Monte Carlo results that indicate that the genus can be used to estimate both the primordial power law quadrupole normalized amplitude $Q_{rms-PS}$ and index $n$, although the pure dependence of the genus amplitude on $n$ is complicated by noise in the data.

   The Differential Microwave Radiometer (DMR) experiment is designed to map the microwave sky and find fluctuations of cosmological origin. For the $7°$ angular scales observed by the DMR, structure is superhorizon size so the spectral and statistical features of the primordial perturbations are preserved (Peebles 1980). We study the DMR maps at 31.5, 53, and 90 GHz and a map with reduced galactic emission formed by removing a model of the sychrotron and dust emissions and using the 31.5 GHz maps as the free-free emission model (see e.g. the subtraction technique of Bennett et al. 1992) and use the above strategies to test the hypothesis that the data are characterized by random-phase initial density perturbations drawn from a Gaussian distribution. Non-Gaussian features are model-specific; we have yet to test the consistency of specific non-Gaussian models with the DMR data. The 2 channels, A and B, at each frequency are combined to form 'sum' (A+B)/2 and 'difference' (A-B)/2 maps. In principle, the sum maps are a combination of Gaussian-distributed instrument noise and CMB anisotropy signal, while the difference maps should consist only of instrument noise. In this paper, we consider only the 4016 pixels corresponding to Galactic latitudes $|b| > 20°$ from which a fitted mean and dipole have been removed. We compare these maps with Monte Carlo realizations of Gaussian power-law models. Each realization consists of the simulation of CMB sky maps using the DMR filter function described in Wright et al. (1994), noise appropriate to the A and B channels at the frequency of interest, reproducing the sky sampling in detail, and then formation of the combination sum and difference maps. Both the DMR and simulated maps are smoothed using Gaussians of varying full-width-at-half-maximums (FWHMs) to sample structure on progressively larger angular scales.

The presence of bright sources in the data could provide a non-Gaussian component to the maps and affect the outcome of statistical and topological tests of Gaussianity. Bennett et al. (1993) give limits to source contributions to the DMR sky maps: it is unlikely that the conclusions of this paper can be affected.

## 2    Statistics of the DMR maps

We consider the properties of various moments of the map temperature distribution: the second moment, $\langle T^2 \rangle$, third moment, $\langle T^3 \rangle$, and fourth moment, $\langle T^4 \rangle$. We neglect the first moment, $\langle T \rangle$, since the mean for any map is set to zero for the region considered (i.e. $|b| > 20°$). We also consider the Kolmogorov-Smirnoff statistic between the sum and difference maps. All of the statistical quantities are evaluated as a function of smoothing angle.

### 2.1    The statistics of noiseless CMB skies

Scaramella & Vittorio (1991) showed that CMB fluctuations observed by a large antenna beam may not be Gaussianly distributed, even if the initial density fluctuations are. We have repeated these calculations for 3000 Harrison-Zel'dovich skies. Our results are consistent with those of Scaramella & Vittorio. This effect is a result of the dominance of the temperature fluctuations by the low order multipoles, which are the least ergodic (Abott & Wise 1984). Figure 1 shows the rms, $\langle T^3 \rangle$ and $\langle T^4 \rangle$ distributions from our simulations. The mean value for $\langle T^3 \rangle$ is consistent with zero as expected for Gaussian fluctuations, but $\langle T^3 \rangle$ has a broad distribution. As pointed out by Graham et al. (1993), it seems unlikely that stringent limits could be placed on cosmological models from this statistic. The mean value of $\langle T^4 \rangle$ is less than expected for pure Gaussian behavior.

## 2.2 THE STATISTICS OF PURE NOISE SKIES

The noise per pixel is well represented by a Gaussian with standard deviation $\sigma_{obs}/\sqrt{N_i}$, where $N_i$ is the number of observations of sky position pixel i and $\sigma_{obs}$ is the instrument noise per observation (Smoot et al. 1990). The DMR sky maps have non-uniform sky coverage, with more observations towards the ecliptic poles, and fewer near the ecliptic plane. A direct consequence of this sky coverage is a departure from Gaussian behavior of pure noise maps. The noise map temperature sample is drawn from Gaussian distributions with variance differing by a factor of four (pixels with $N_i$ ranging from 8900 to 42000), creating a non-Gaussian distribution. The noise map temperature distribution is expected to show more outliers (positive kurtosis) than a Gaussian distribution of identical variance. Figure 2 compares the temperature distribution of 3000 realizations of pure noise to an exact Gaussian distribution. The variance alone is insufficient to distinguish Gaussian from non-Gaussian behavior. $\langle T^3 \rangle$ is consistent with zero as expected, since no sign is favored in the sum of Gaussians each with a zero mean. $\langle T^4 \rangle$ is shifted in the positive direction relative to the value expected for Gaussian statistics as expected for a sample drawn from Gaussians with different variances.

## 2.3 STATISTICS OF NOISY CMB SKY MAPS

### 2.3.1 Weighting schemes

Since both noiseless CMB and pure noise maps demonstrate non-Gaussian behavior from Gaussian initial conditions, it is also likely that noisy CMB maps deviate from Gaussian statistics. For noisy data, a weighting scheme should be employed to minimize the contribution from the most noisy pixels. The most appropriate weighting schemes depend on both the noise properties and unknown CMB temperature at each pixel.

The *COBE* DMR temperature sky maps are comprised of the underlying cosmic temperature distribution $T_i$ plus a noise term $n_i$ for each pixel i. We assume nothing about the statistical properties of $T_i$, since that is what we are attempting to determine. The measured temperature value at each pixel i, is $T_{obs,i} = T_i + n_i$. In what follows, the $T_{obs,i}$ have had a mean and dipole removed (using weighting by $N_i$) before the moments are evaluated. We define the weighted quantity to order p:

$$\langle T^p \rangle \equiv \frac{\sum_i T_i^p w_i}{\sum_i w_i},$$

| Quantity $\langle T^p \rangle$ | $< n_i^p >$ | $w_i = 1/\text{Variance}$ |
|---|---|---|
| $\langle T \rangle \equiv$ Mean | 0 | $N_i/\sigma_{obs}^2$ |
| $\langle T^2 \rangle \equiv$ Variance | $\sigma_{obs}^2/N_i$ | $N_i^2/2\sigma_{obs}^4$ |
| $\langle T^3 \rangle \equiv$ Third Moment | 0 | $N_i^3/15\sigma_{obs}^6$ |
| $\langle T^4 \rangle \equiv$ Fourth Moment | $3\sigma_{obs}^4/N_i^2$ | $N_i^4/96\sigma_{obs}^8$ |

Table 1: Weighting, $w_i$, for a map dominated by Gaussian noise with variance $\sigma_{obs}^2/N_i$.

where $w_i$ are the weights. For any Gaussianly-distributed quantity, the correct statistical weight is given by the inverse variance of the quantity. If the map is dominated by noise, then the best weights for $\langle T^p \rangle$ are proportional to $N^p$ as shown in Table 1. However, when there is comparable or greater signal than noise, the optimum weighting is significantly different. While analytic expressions can be found for the optimum weighting, they depend on the unobserved temperatures $T_i$ instead of the observed values $T_{obs,i}$. The weighting schemes are sensitive to the signal-to-noise ratio as the smoothing FWHM varies. We find that the most reliable way to handle the issue is to compare the data with detailed Monte Carlo simulations.

### 2.3.2  Bias terms

The presence of noise in the data generates additional 'bias' terms. The relationship of the observed map temperature moments to the desired CMB temperature moments are:

$$\langle T_{obs} \rangle = \langle T \rangle + \langle n \rangle$$

$$\langle T_{obs}^2 \rangle = \langle T^2 \rangle + 2\langle Tn \rangle + \langle n^2 \rangle$$

$$\langle T_{obs}^3 \rangle = \langle T^3 \rangle + 3\langle T^2 n \rangle + 3\langle Tn^2 \rangle + \langle n^3 \rangle$$

$$\langle T_{obs}^4 \rangle = \langle T^4 \rangle + 4\langle T^3 n \rangle + 6\langle T^2 n^2 \rangle + 4\langle Tn^3 \rangle + \langle n^4 \rangle$$

The terms containing odd powers of the noise vanish asymptotically when summed over many pixels. There still remain bias terms involving even powers of the noise. For the sample variance in the sum map, $\langle T^2 \rangle$,

$$\langle T^2 \rangle \big|_{\text{Sum}} = \frac{1}{\Gamma} \sum_i (T_i + n_i)^2 \, w_i$$

where $\Gamma = \sum_i w_i$. We can estimate the noise term by using the difference map,

$$\langle T^2 \rangle \big|_{\text{Diff}} = \frac{1}{\Gamma} \sum_i n_i^2 w_i$$

allowing an estimate of the (noiseless) sky variance as the difference

$$\langle T^2 \rangle \big|_{\text{Sky}} = \langle T^2 \rangle \big|_{\text{Sum}} - \langle T^2 \rangle \big|_{\text{Diff}}$$

$$\langle \text{T}^2 \rangle \big|_{\text{Sky}} = \frac{1}{\Gamma} \sum_i \left( \text{T}_i^2 + 2\text{T}_i \text{n}_i + \text{n}_i^2 \right) \text{w}_i - \frac{1}{\Gamma} \sum_i \text{n}_i^2 \text{w}_i \sim \frac{1}{\Gamma} \sum_i \text{T}_i^2 \text{w}_i$$

In the following analysis we drop all terms that asymptotically approach zero, whether or not they must strictly go to zero in a single map. The difference of the sum and difference map variance is an estimator of the sky variance free from additional bias terms. We have simulated the sky rms to test this estimator against the input sky rms using a number of weighting schemes, from unit pixel weighting up to $\text{N}_i^4$ (for consistency with later calculations). Figure 3 compares the noisy estimator to the input noiseless sky. Weights proportional to $\text{N}_i^2$ produce the smallest scatter about the ensemble average and are optimal in the sense of providing the greatest statistical power. The scatter in the unit weighting case is not much greater than in the $\text{N}_i^2$ weighting case: although $\text{N}_i^2$ weighting minimizes the scatter due to noise, it increases the observed spread due to cosmic variance. However, the weighted sum

$$\frac{1}{\Gamma} \sum_i \text{T}_i^2 \text{w}_i$$

is not the same as the noiseless sky variance

$$\frac{1}{\text{M}} \sum_i \text{T}_i^2$$

summed over M pixels except in the case of unit weighting, $\text{w}_i = 1$. Although the weighted estimators successfully recover the *weighted* CMB variance, this statistic is experiment-specific since the weights $\text{N}_i^\text{p}$ depend on the observation pattern on the sky. Weighted estimators must be calibrated using Monte Carlo simulations to extract useful cosmological information from noisy sky maps. The unweighted estimator, although of lesser statistical power, is more easily compared between experiments.

For the third and fourth moments, the situation is more complex, since additional terms are present even if we try to combine the values from the sum and difference maps. For the third moments, $\langle \text{T}^3 \rangle$,

$$\langle \text{T}_{\text{obs}}^3 \rangle \big|_{\text{Sum}} = \frac{1}{\Gamma} \sum_i \left( \text{T}_i + \text{n}_i \right)^3 \text{w}_i$$

$$\langle \text{T}_{\text{obs}}^3 \rangle \big|_{\text{Diff}} = \frac{1}{\Gamma} \sum_i \text{n}_i^3 \text{w}_i.$$

The noise term $\langle \text{n}^3 \rangle$ is approximately zero when averaged over the map. An estimator of the sky third moment need only consider the ensemble average of the sum map value.

$$\langle \text{T}_{\text{obs}}^3 \rangle \big|_{\text{Sum}} \sim \frac{1}{\Gamma} \sum_i \left( \text{T}_i^3 + 3\text{T}_i \text{n}_i^2 \right) \text{w}_i$$

The last term introduces a bias, which can be written as $3\sigma_{\text{obs}}^2 / \Gamma \sum_i \text{T}_i \text{w}_i / \text{N}_i$. This term represents a 'beating' of the temperature distribution $\text{T}_i$ with the observation field $N_i$. Note that *if* we choose our weighting $w_i$ to be equal to $N_i^2$, then the bias term becomes the sum $\frac{1}{\Gamma} \sum_i T_i N_i$, which must be zero, since this is the previously subtracted mean. This choice of weight eliminates the bias, but neither minimizes the scatter about the ensemble average nor reproduces the unweighted statistic for a noiseless sky. Our Monte Carlo realizations show that *no* weighting scheme can both remove the bias term and recover the underlying CMB $\langle T^3 \rangle$ value (Figures 3c, d). $N_i^3$ weighting minimizes the spread for noise dominated maps; however, $N_i^2$ weighting minimizes the total spread in maps with noise and power spectrum cosmic variance and is optimal for the Monte Carlo comparison of noisy data to theoretical models.

As commented in a number of recent articles, the $\langle T^3 \rangle$ statistic has such a broad distribution even in the absence of noise that statements about the DMR values in terms of any cosmological models will have little statistical power. The situation appears to be worse than estimated analytically by Srednicki (1993) who gives a value of 1.3 $Q_{rms-PS}^3$ for the standard deviation of the $\langle T^3 \rangle$ distribution for a sky with the quadrupole removed (including a correction factor to allow for incomplete sky coverage). Our simulations of noiseless $n = 1$ CMB skies with no quadrupole subtraction gives a value almost 50% higher. Since the noise standard deviation in $\langle T^3 \rangle$ is at least a factor of 10 higher (without additional smoothing), the difficulty in the use of this statistic is further emphasized.

For the 4th moments, $\langle T^4 \rangle$, we have

$$\langle T_{obs}^4 \rangle \big|_{\text{Sum}} = \frac{1}{\Gamma} \sum_i (T_i + n_i)^4 w_i$$

and

$$\langle T_{obs}^4 \rangle \big|_{\text{Diff}} = \frac{1}{\Gamma} \sum_i n_i^4 w_i.$$

Since $\langle n^4 \rangle$ does not have a zero expectation value, we need to take the difference between the sum and difference statistics to estimate the sky value,

$$\langle T_{obs}^4 \rangle \big|_{\text{Sum}} - \langle T_{obs}^4 \rangle \big|_{\text{Diff}} = \frac{1}{\Gamma} \sum_i \left( T_i^4 + 4T_i^3 n_i + 6T_i^2 n_i^2 + 4T_i n_i^3 \right) w_i,$$

Eliminating terms in $n_i$ and $n_i^3$ gives

$$\langle T_{obs}^4 \rangle \big|_{\text{Sum}} - \langle T_{obs}^4 \rangle \big|_{\text{Diff}} = \frac{1}{\Gamma} \sum_i \left( T_i^4 + 6T_i^2 n_i^2 \right) w_i.$$

There is a bias that can be written as $6\sigma_{obs}^2/\Gamma \sum_i T_i^2 w_i/N_i$. In principle, this term can be made to depend on the sky variance. For example, if we consider that the best weight to use to calculate the variance is $N_i^2$ then by defining the weight $w_i$ in the above equation to be $N_i^3$ the bias term will be given by $6\sigma_{obs}^2 \text{Var}(\text{sky}) \sum_i N_i^2 / \sum_j N_j^3$ and may be subtracted to estimate $\langle T^4 \rangle$. The situation is better than in the case of the third moment since the spread in the realizations is much smaller. The best weight in the sense of minimizing the spread in the simulated values is indeed $N_i^3$ for the combination of cosmic variance and noise, while $N_i^4$ is better for noise-dominated maps.

We conclude that only in the case of the sky variance can noisy maps provide an unbiased estimate of the true sky value. Higher-order moments must be compared to specific CMB models through Monte Carlo simulations. The strictest statistical limits can be placed on the comparisons by using weights which minimize the spread in the simulated data.

### 2.4   DMR sky map statistics

The Kolmogorov-Smirnoff statistic provides a simple test of the hypothesis that two samples are drawn from the same parent distribution. Figure 4 shows the Kolmogorov-Smirnoff statistic for the 53 GHz sum and difference maps as a function of smoothing angle. For small smoothing angles, the sum map is dominated by the noise contribution and the null hypothesis is accepted at 44% confidence. At smoothing angles > 4° the CMB fluctuations become important relative to the noise and the null hypothesis is rejected. We consider whether the CMB fluctuations represent Gaussian initial conditions. Figure 5 shows the weighted moments of the 53 GHz sum and difference maps as a function of smoothing angle, compared to 1000 simulations of Gaussian power-law models with $n = 1$, $Q_{rms-PS} = 17$ $\mu$K and DMR instrumental noise. Contributions to $\langle T^2 \rangle$, $\langle T^3 \rangle$, and $\langle T^4 \rangle$ from the Galaxy (Bennett et al. 1992) and systematic artifacts (Kogut et al. 1992) are negligible for $|b| > 20°$.

Wright et al. (1994) have used the sky variance at 10° effective smoothing (7° FWHM smoothing of the pixelized maps) to derive limits on the quadrupole parameter $Q_{rms-PS}$ for Gaussian models with $n = 1$. We use the higher-order moments and the noise properties deduced from the difference maps to determine both $n$ and $Q_{rms-PS}$ via a least-squares minimization of the $\chi^2$ statistic

$$\chi^2 = \sum_{i, j} (T^p_{obs} - \langle T^p_{sim} \rangle)_i \, \mathbf{M}^{-1}_{ij} \, (T^p_{obs} - \langle T^p_{sim} \rangle)_j$$

where the indices i and j refer to the smoothing angle, $\mathbf{M}$ is the covariance matrix from the simulated data, and the moments are weighted by $N^p$ to minimize the scatter from the instrument noise. The simulations include cosmic variance. Figure 6 shows the confidence interval from fitting the rms temperature variations at various smoothing angles to a grid of $Q_{rms-PS}$ and $n$ giving $Q_{rms-PS} = 13.2 \pm 2.5$ $\mu$K and a power law index $n = 1.7^{+0.3}_{-0.6}$. The chi-squared contours are very elliptical because of the high degree of correlation between $Q_{rms-PS}$ and $n$ thus the relation $Q_{rms-PS} = (16.7 \pm 2.6)$ - $(4.1 \pm 1.3)(n - 1)$ $\mu$K. Combining the various moments, $\langle T^p \rangle$, does not improve the result significantly and as noted the higher moments are more susceptible to biases caused by the noise. When $n$ is forced to unity, we find $Q_{rms-PS} = 18.5 \pm 3.2$. When $n$ is forced to unity and we use only the 10° -smoothed data we recover $Q_{rms-PS} = 17.5 \pm 2.3$ in agreement with Wright et al. (1994).

## 3 The genus

The statistical properties of CMB fluctuations may be characterized by the excursion regions enclosed by isotemperature contours. The genus is the total curvature of such isotemperature contours. The genus per unit solid angle is a locally invariant quantity in the sense that incomplete sky coverage and coordinate transformations leave the quantity intact. This is of importance for the DMR sky maps, since the genus will be insensitive to the Galactic cut angle imposed on the data. The genus can be roughly defined as the number of isolated high temperature regions (hot spots) minus the number of isolated low temperature regions (cold spots). Bond & Efstathiou (1987) give approximate analytic formulae for the expected number of hot and cold spots in a realization of a Gaussian random field. Gott et al. (1990) have provided a full treatment of this quantity for CMB skies, thus we only summarize for the application of the genus to the DMR sky maps.

For a random Gaussian field with a correlation function $C(\theta)$ and rms temperature fluctuation $\sigma$, the expectation value of the total genus on a sphere is,

$$G_s = 4\pi g + \mathrm{erfc}(\nu/\sqrt{2})$$

where erfc is the complementary error function, $\nu$ is the threshold above the mean in units of $\sigma$ and g, the mean genus per unit area, is

$$g = \frac{1}{(2\pi)^{3/2}\theta_c^2}\nu e^{-\nu^2/2}$$

The correlation angle $\theta_c$ is related to the correlation function $C(\theta)$ according to $\theta_c^{-2} = -\frac{1}{C(0)}\frac{d^2C(\theta)}{d\theta^2}\Big|_{\theta=0}$ (Adler 1981).

The genus provides useful information about a set of data in two ways: the shape tests the Gaussian nature of the data, while the amplitude determines properties of the model being tested. Figure 7 shows the mean genus values for Monte Carlo realizations of the sky as seen by the DMR beam for different values of $n$. For Gaussian fields with power law initial density fluctuations, the amplitude of the genus is a function of the power law spectral index $n$. The characteristic shape of the genus curve for Gaussian fluctuations can also be seen. This can be understood as follows: At low temperature thresholds, the isotemperature contours surround cold spots on the map and the total curvature is negative. At high temperature thresholds the contours surround hot spots and the curvature is positive. Near the mean, the total curvature is close to zero. Note that this is an unbiased estimator of Gaussianity. The large DMR antenna beam does not cause the genus curve to deviate from the shape expected for Gaussian fluctuations, unlike the temperature moments which can be significantly modified. This can be seen from the self-similarity between the curves for a noiseless CMB and pure noise realization. In the presence of Gaussian noise, the genus still has a characteristic shape and trend with spectral index, but the amplitude now depends on the signal-to-noise ratio of the map. Since the signal can be characterised by the rms-quadrupole-normalized amplitude $Q_{rms-PS}$, the genus represents an independent method of estimating $Q_{rms-PS}$ and $n$ for the DMR sky maps.

We determine the genus in two independent ways: using the complete definition in terms of curvature, and as the difference in the number of hot and cold spots. Both methods give consistent results. The genus is evaluated for the DMR sum and difference sky maps as a function of both threshold $\nu$ and smoothing angle, from 0 to 20° FWHM. Figure 8 shows the genus curve for the 53 GHz maps for 5° smoothing and 20° galaxy cut.

We test the genus of the DMR sum and difference maps against Monte Carlo simulations consisting of instrument noise plus a power-law model of CMB anisotropy characterized by amplitude $Q_{rms-PS}$ and index $n$ or instrument noise alone, allowing for the different signal-to-noise ratios expected for the DMR maps at different frequencies. We define a chi-squared statistic

$$\chi^2 = \sum_{j,i} \sum_{l,k} \left( (G_{ij} - \bar{G}_{ij}) M_{ij,kl}^{-1} (G_{kl} - \bar{G}_{kl}) \right)$$

where $G_{ij}$ is the genus for the $i^{th}$ bin in contouring threshold and the $j^{th}$ smoothing angle, $\bar{G}_{ij}$ is the mean of all simulations for that bin, and $M_{ij,kl}^{-1}$ is the covariance between bins found from simulations. We use contour levels ranging from -2.5 to +2.5 in steps of 0.5 and smoothing angle FWHM of 0, 5, 10, 15, and 20 degrees for a total of 55 bins. We convert the chi-squared to a probability, $P(\chi^2)$, defined as the fraction of simulations at identical threshold and smoothing angles whose $\chi^2$ values were larger than the DMR value. If the DMR data are near the median of the $\chi^2$ distribution then $P \approx 0.5$. Table 2 summarizes the results for the 3 DMR frequencies and the reduced-galaxy maps. The DMR 53 GHz data (P=0.5) are consistent with a superposition of noise and Gaussian, scale-invariant CMB. The DMR 31 GHz and 90 GHz data are qualitatively similar to the 53 GHz data but have poorer probabilities. Only 3% of the simulations have a $\chi^2$ as large as the 90 GHz data and only 0.3% as large as the 31 GHz data. From the shift in fitted power-law parameters to more power at large scales, it seems likely that the Galactic foreground adds a significant bias.

Figure 9 shows the confidence intervals for the 53 GHz sum map genus. It is evident that the genus statistic is more sensitive to $Q_{rms-PS}$ than to $n$ for the DMR first year sky maps. For the noise levels of the first year of data, the fitted $Q_{rms-PS}$ and $n$ are highly correlated and one obtains an acceptable statistical fit for a range of parameters along the line

$Q_{rms-PS} = (15.7 \pm 2.2)$ - $(6.6 \pm 0.3)(n - 1)$ $\mu K$.

The best fitted values for the 53 GHz sum map are $Q_{rms-PS} = 12$ $\mu K$ and $n = 1.7$, but it is not significantly better than the fitted value with $n$ forced to 1, $Q_{rms-PS} = 15.7 \pm 2.2$ $\mu K$.

The degeneracy between $Q_{rms-PS}$ and $n$ results from cosmic variance and instrument noise, which has an effective index $n \approx 3$. Figure 10 shows the fitted $Q_{rms-PS}$ and $n$ from 110 simulations of noisy maps drawn from a parent population with $Q_{rms-PS} = 17$ $\mu K$, $n=1$, and noise equivalent to the 53 GHz sum map. We recover $Q_{rms-PS} = (16.5 \pm 2.2)$ - $(6.6 \pm 0.3)(n-1)$ $\mu K$. The fitted values for $n = 1$ differ by 0.5 $\mu K$ from the input. All the genus results in this paper have been adjusted for this bias. A similar analysis with noise levels equivalent to four years of data indicates reasonable decoupling between the fitted $Q_{rms-PS}$ and $n$.

| Map | Fitted $Q_{rms-PS}$ $\mu K$ | Fitted $n$ | Maximum Probability | $Q_{rms-PS}(n=1)$ $\mu K$ | Probability |
|---|---|---|---|---|---|
| 31 (A+B)/2 | 28 | 0.0 | 0.3% | $25.9 \pm 6.8$ | $< 0.3\%$ |
| 53 (A+B)/2 | 12 | 1.7 | 53% | $15.7 \pm 2.2$ | 50% |
| 90 (A+B)/2 | 13 | 1.5 | 3.1% | $18.3 \pm 3.9$ | 2.8% |
| Reduced Galaxy | 13 | 1.5 | 9.1% | $14.7 \pm 4.2$ | 6.6% |
| | | | | | |
| 31 (A-B)/2 | 18 | 1.0 | 6.7% | $20 \pm 7$ | 6.7% |
| 53 (A-B)/2 | 4 | 1.5 | 6.5% | $4 \pm 3$ | 2.3% |
| 90 (A-B)/2 | 0 | 0.5 | 9.1% | $0^{+2}_{-0}$ | 4.8% |
| RG (A-B)/2 | 7 | 1.5 | 14% | $6 \pm 5$ | 14% |

Table 2: Best fitted values for $Q_{rms-PS}$ and $n$ using the genus.

## 4   Discussion

We have demonstrated that some of the statistical quantities used to distinguish Gaussian from non-Gaussian behavior are not reliable since the presence of noise in the real data leads to biases in the results. $\langle T^3 \rangle$ has such a broad distribution due to cosmic variance that even in the absence of noise it is not a useful statistic. $\langle T^4 \rangle$ is of somewhat more use, but the 'true' sky value is not recoverable. It should not be too surprising that these statistics are of limited use, since they do not utilize any information on spatial correlations in the data, and are themselves invariant under spatial scrambling of the data at their initial resolution. Smoothing the data does provides sensitivity to extended sky structure. More useful 3-point statistics make use of spatial correlations in the data. Hinshaw et al. (1994) describe the 3-point correlation function in detail, while Graham et al. (1993) propose an alternative 3-point statistic. The genus is a more useful quantity, and has the advantage that it also provides an estimate of the power law amplitude and spectral index.

The temperature moments and the genus of the DMR data are consistent with a superposition of instrument noise and CMB anisotropy with a power law amplitude from a random-phase Gaussian distribution. This result is supported by the analysis of the 3-point correlation function in Hinshaw et al. (1994). We use both the genus and the temperature moments to estimate the parameters $Q_{rms-PS}$ and $n$. With

$Q_{rms-PS}$ and $n$ both free, we recover fitted values

$$Q_{rms-PS} = 13.2 \pm 2.5 \ \mu K \qquad n = 1.7^{+0.3}_{-0.6}$$

from the rms evaluated at different smoothing angles, and

$$Q_{rms-PS} = (15.7 \pm 2.2) \text{ -} (6.6 \pm 0.3)(n \text{ - } 1) \ \mu K$$

from the genus. The fitted $Q_{rms-PS}$ and $n$ are highly correlated. With $n$ forced to unity we recover values

$$Q_{rms-PS} = 18.5 \pm 3.2 \qquad \text{(rms vs smoothing angle)}$$
$$Q_{rms-PS} = 15.7 \pm 2.2 \qquad \text{(genus)}.$$

These estimates are consistent with previous evaluations of the DMR data, $Q_{rms-PS}$ = $17 \pm 5 \ \mu K$, $n = 1.1 \pm 0.5$ from fits to the 2-point correlation function (Smoot et al. 1992, Wright et al. 1992) and $Q_{rms-PS} = 17.1 \pm 2.9 \ \mu K$ from forced $n = 1$ fitted to the sky variance at 10 degrees effective smoothing (Wright et al. 1994).

The DMR data are consistent with Gaussian initial density perturbations. The DMR data are inconsistent with strongly non-Gaussian distributions, e.g. a top hat and distributions with excessive tails. However, we have yet to test the DMR data against specific non-Gaussian models of structure formation. The power of the tests improves rapidly with increasing signal to noise ratio. For a given statistic $\langle T^p \rangle$, the noise falls with time, t, as $t^{-p/2}$. With four years of data the results will be limited primarily by cosmic variance.

**Acknowledgements:** The *COBE* DMR research is a team effort; we are grateful to all who have contributed. We thank Changbom Park for providing us with his software for calculating the genus. The code has been slightly modified to fit our specific application so that any errors are ours. Most of our final results were based upon the counting of spots but it was reassuring to have both software results agree. We acknowledge the support of the Office of Space Sciences of NASA Headquarters.

**Figure Captions**

**Figure 1**: Statistical distributions for noiseless CMB maps: a) rms or $\langle T^2 \rangle^{0.5}$, b) $\langle T^3 \rangle$, c) $\langle T^4 \rangle$. The dashed lines show the expected values for Gaussian statistics with the input rms. Note the large spread due to cosmic variance and that the mode is displaced from the value expected for Gaussian statistics.

**Figure 2**: Statistical distributions for pure noise maps with DMR 53 GHz sky coverage and rms per observation: a) a histogram of the temperature distribution and a matching Gaussian, b) rms or $\langle T^2 \rangle^{\frac{1}{2}}$, b) $\langle T^3 \rangle$, c) $\langle T^4 \rangle$. The dashed lines show the expected values for Gaussian statistics. Note that the fourth moment, $\langle T^4 \rangle$, is biased to greater values than Gaussian.

**Figure 3**: Statistics of simulated noisy CMB skies compared to simulated noiseless CMB skies. The simulations are for $Q_{rms-PS} = 17$ $\mu$K, $n = 1$, and 53 GHz noise values are used. a) rms, uniform weighting, b) rms, $N_i^2$ weighting, c) $\langle T^3 \rangle$, $N_i^2$ weighting (to eliminate bias), d) $\langle T^3 \rangle$, $N_i^3$ weighting (to minimize noise scatter) e)$\langle T^4 \rangle$, $N_i^3$ weighting (to eliminate bias), f) $\langle T^4 \rangle$, $N_i^4$ weighting (to minimize noise scatter). The gray bands are the unweighted statistics with cosmic variance only - no noise. The points (*) are the mean of the simulations including noise and cosmic variance with 68% CL error bars. Note that the $N_i^{p-1}$ weighting tends to give a narrower error bar due to cosmic variance effects than the $N_i^p$ that is best for noise-dominated maps. Note also the bias in even powers, p, caused by the noise beating with the signal. It is particularly evident in the $\langle T^4 \rangle$ plot at small smoothing angles.

**Figure 4**: Kolmogorov-Smirnoff test that the DMR 53 GHz sum and difference maps are drawn from the same parent distribution. The gray band shows the mean and 68% confidence interval for the Kolmogorov-Smirnoff statistic as a function of smoothing angle, in the case where both maps are drawn from a pure noise distribution. For smoothing angles $> 4°$, the null hypothesis is rejected.

**Figure 5**: DMR 53 GHz sum and difference map statistics as a function of smoothing FWHM compared to simulated skies with $Q_{rms-PS} = 17$ $\mu$K, $n = 1$, and 53 GHz noise values. The comparisons are made with $N^p$ weighting. a) rms, sum map, b) rms, difference map, c) $\langle T^3 \rangle$, sum map, d) $\langle T^3 \rangle$, difference map, e) $\langle T^4 \rangle$, sum map, f) $\langle T^4 \rangle$, difference map. The data are shown as points marked '*'.

**Figure 6**: Confidence contours for the fit in $Q_{rms-PS}$ and $n$ between the DMR 53 GHz $\langle T^2 \rangle$ and Monte Carlo simulations. The contours are for the 68%, 95%, and 99% confidence intervals, determined by the probability distribution of our simulations. The best fit is marked by an asterisk (*).

**Figure 7**: The genus curve for noiseless Gaussian power-law initial fluctuations of different $n$, each evaluated at 5° FWHM smoothing. The genus from instrument noise with the DMR observation pattern is also shown, reduced by a factor 5 to fit on the same scale.

**Figure 8**: 53 GHz genus curves for 5° smoothing: a) sum and difference map values compared to simulated sum maps with $Q_{rms-PS} = 15.7$ $\mu$K, $n = 1$, and 53 GHz noise properties, b) sum and difference map values compared to simulated difference

maps with 53 GHz noise properties. The gray band is the 68% confidence interval for the simulations. The sum map is consistent with Gaussian power-law density perturbations plus noise and is inconsistent with pure instrument noise.

**Figure 9**: Confidence contours for the fit for $Q_{rms-PS}$ and $n$ between the DMR 53 GHz genus curves and simulated genus curves. The contours are for the 68%, 95%, and 99% confidence intervals. The best fit is marked by an asterisk (*).

**Figure 10**: Scatter plot of best-fitted $Q_{rms-PS}$ and $n$ using the genus from 110 realizations of noisy maps drawn from a parent population with $Q = 17$ $\mu$K and $n = 1$. The fitted slope is $Q_{rms-PS}(n) = (16.5 \pm 0.2)$ - $(6.6 \pm 0.3)(n - 1)$ $\mu$K (solid line). We also show the functional form $Q_{rms-PS}(n) = 17 \exp(0.46(1-n))$ using the slope suggested by Seljak and Bertschinger (1993) for the relation between $Q_{rms-PS}$ and $n$ evaluated from the 2-point correlation function, normalized to this simulation.

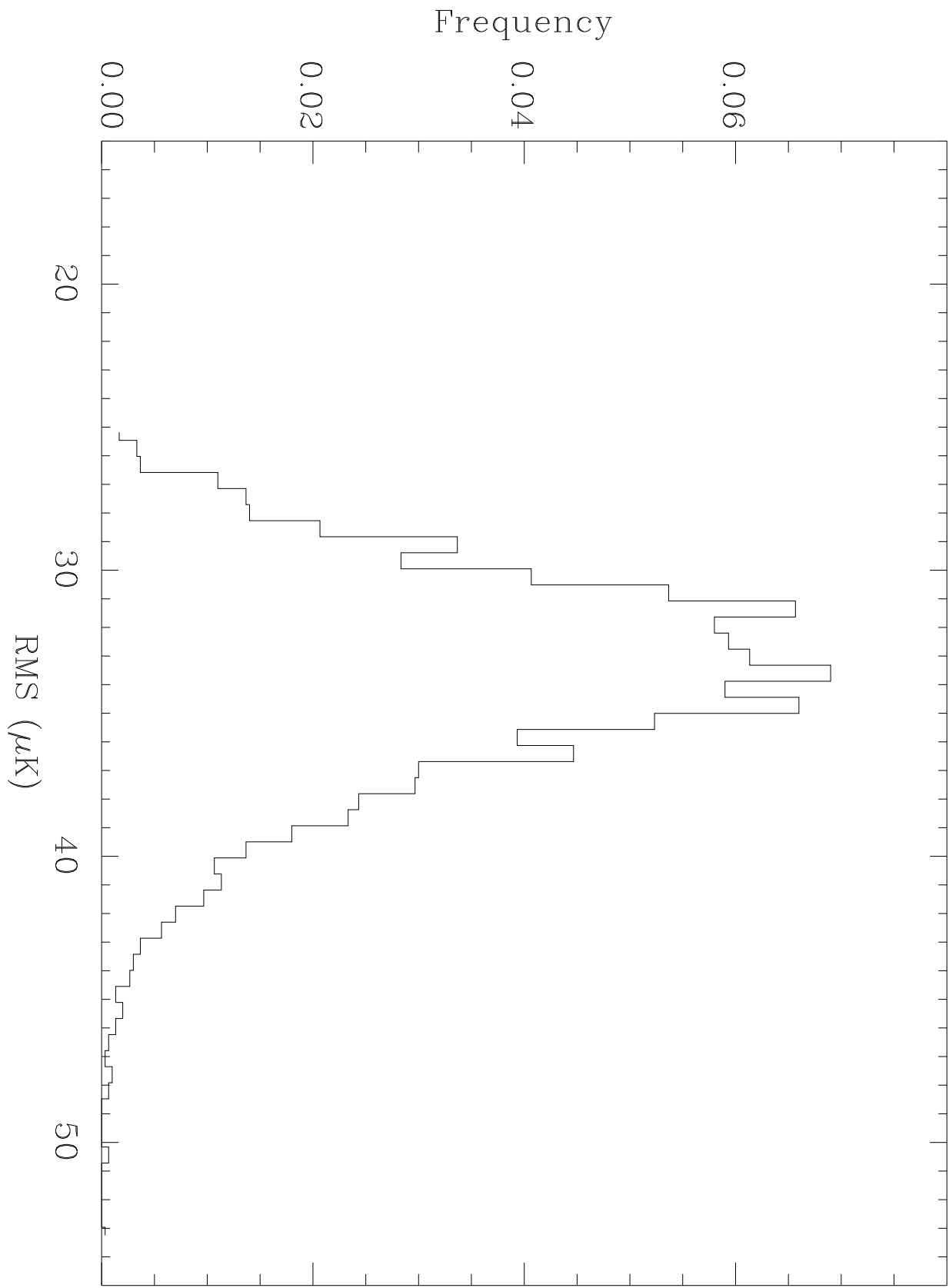

Figure 1a

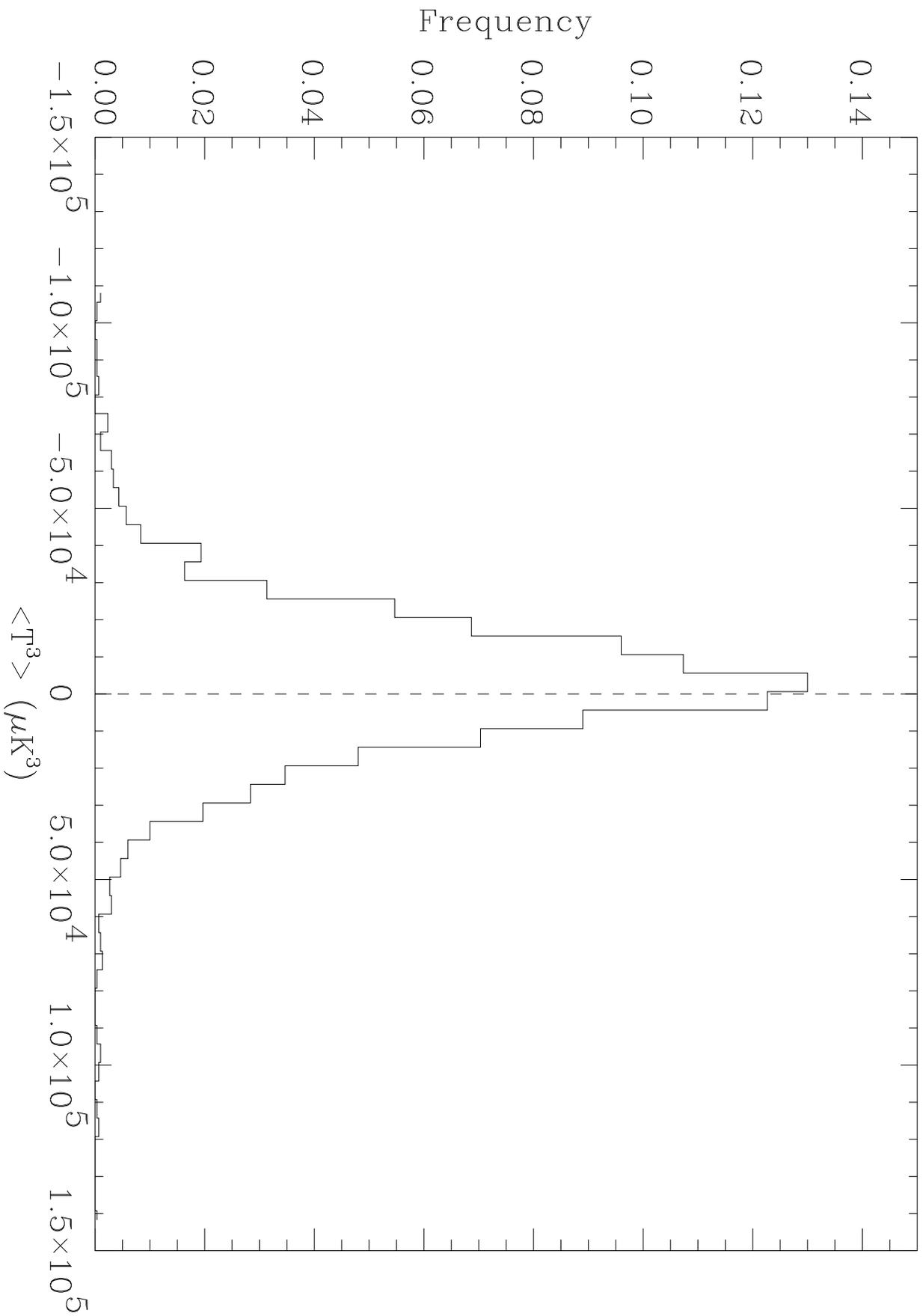

Figure 1b

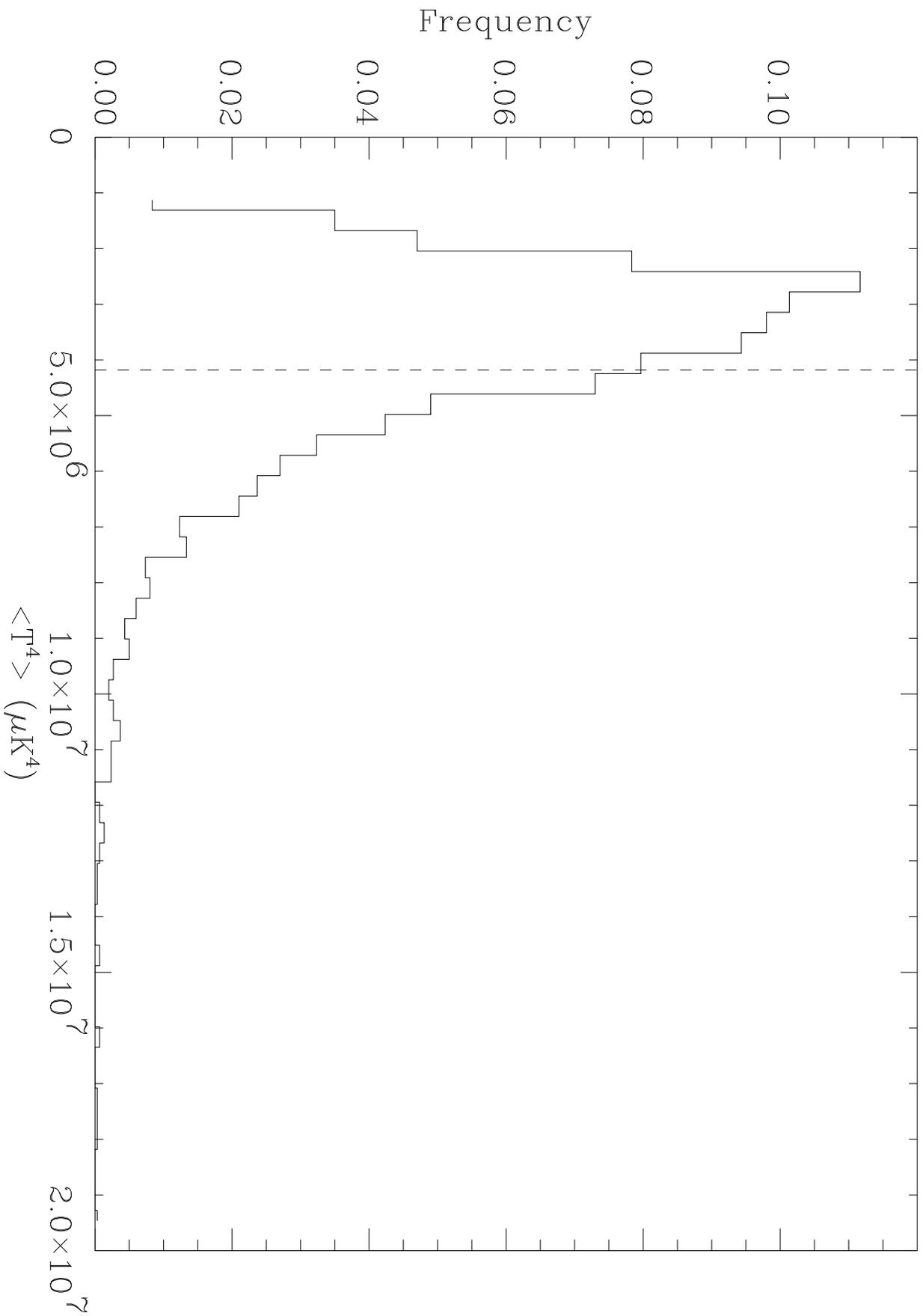

Figure 1c

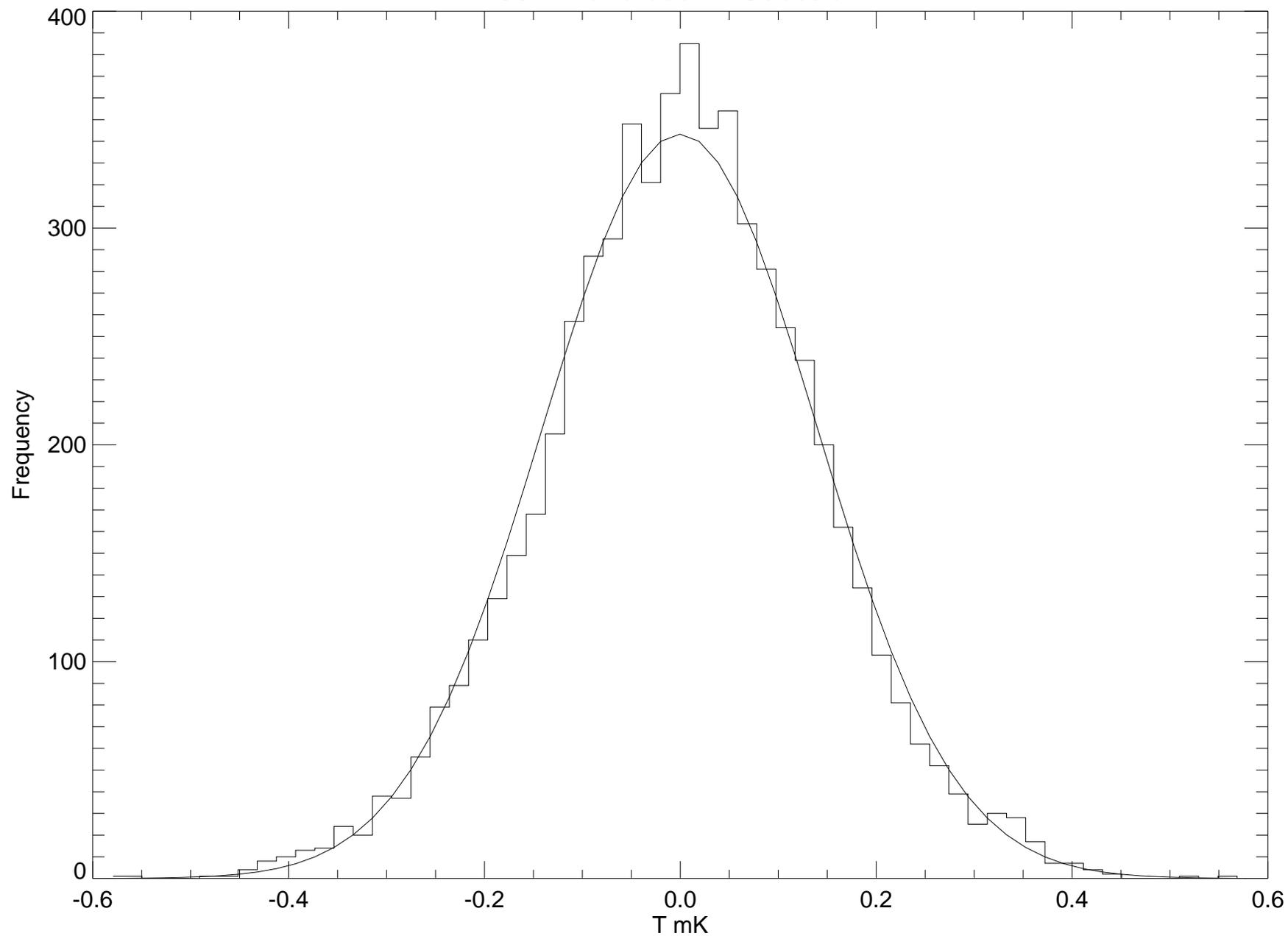

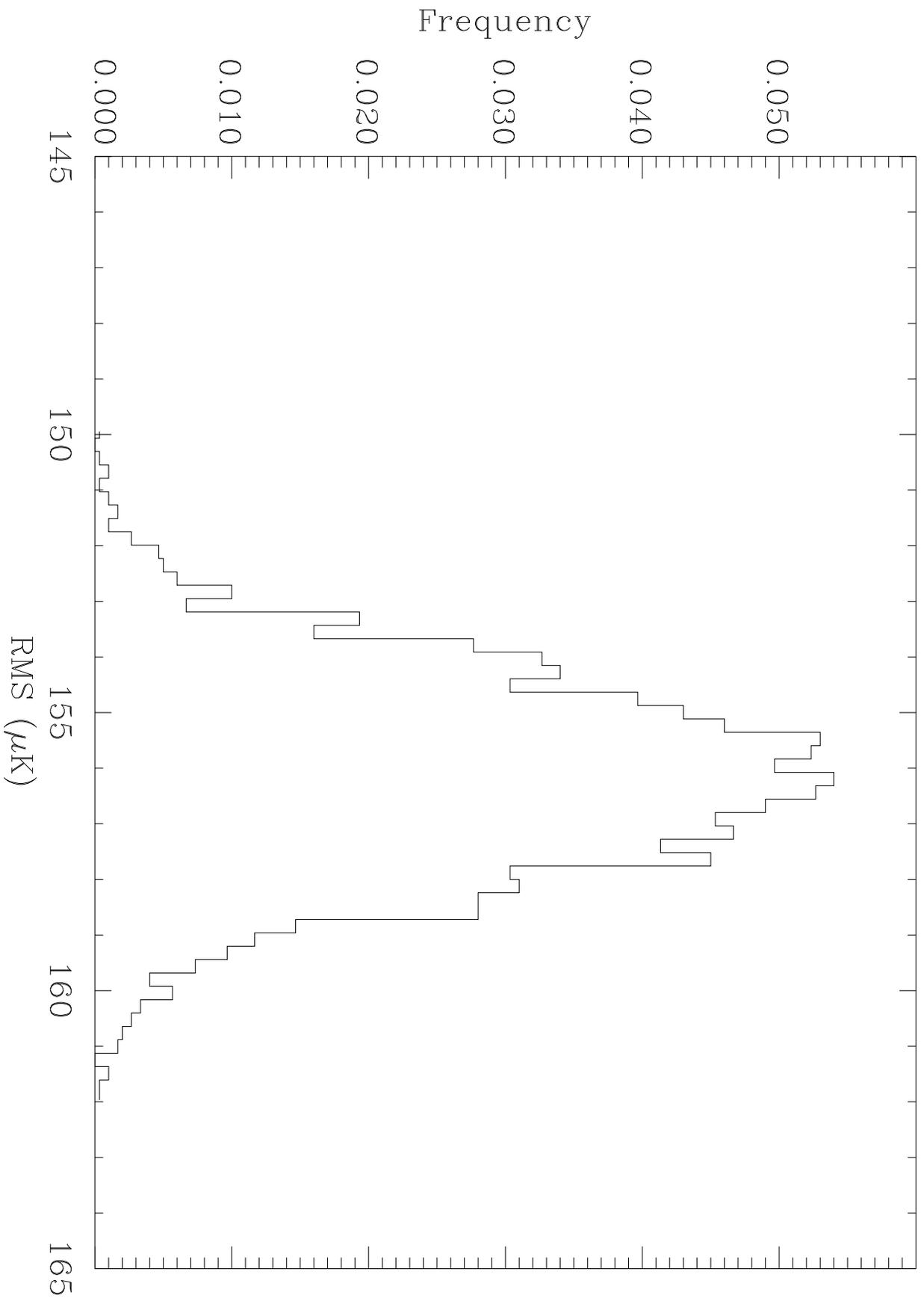

Figure 2b

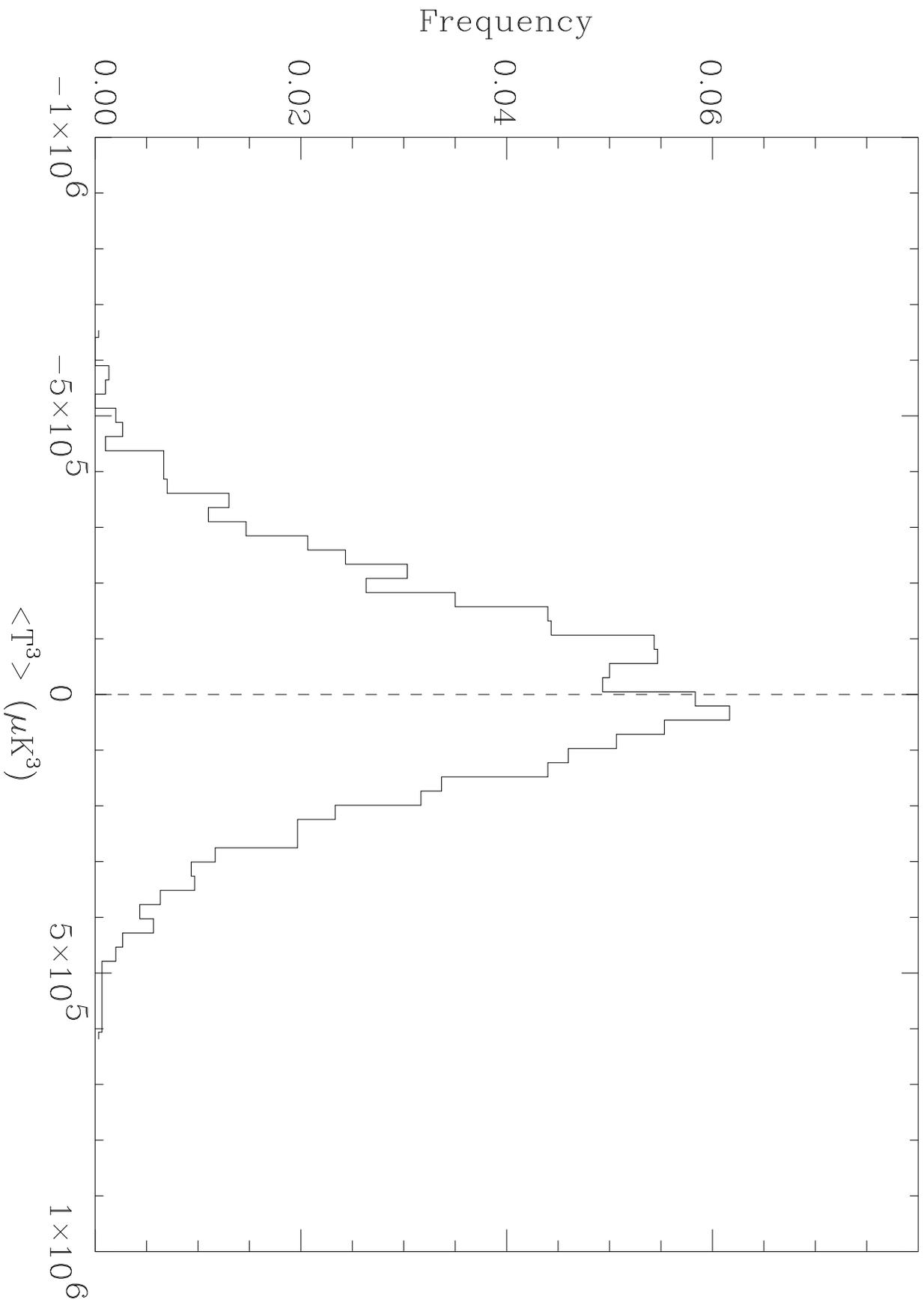

Figure 2c

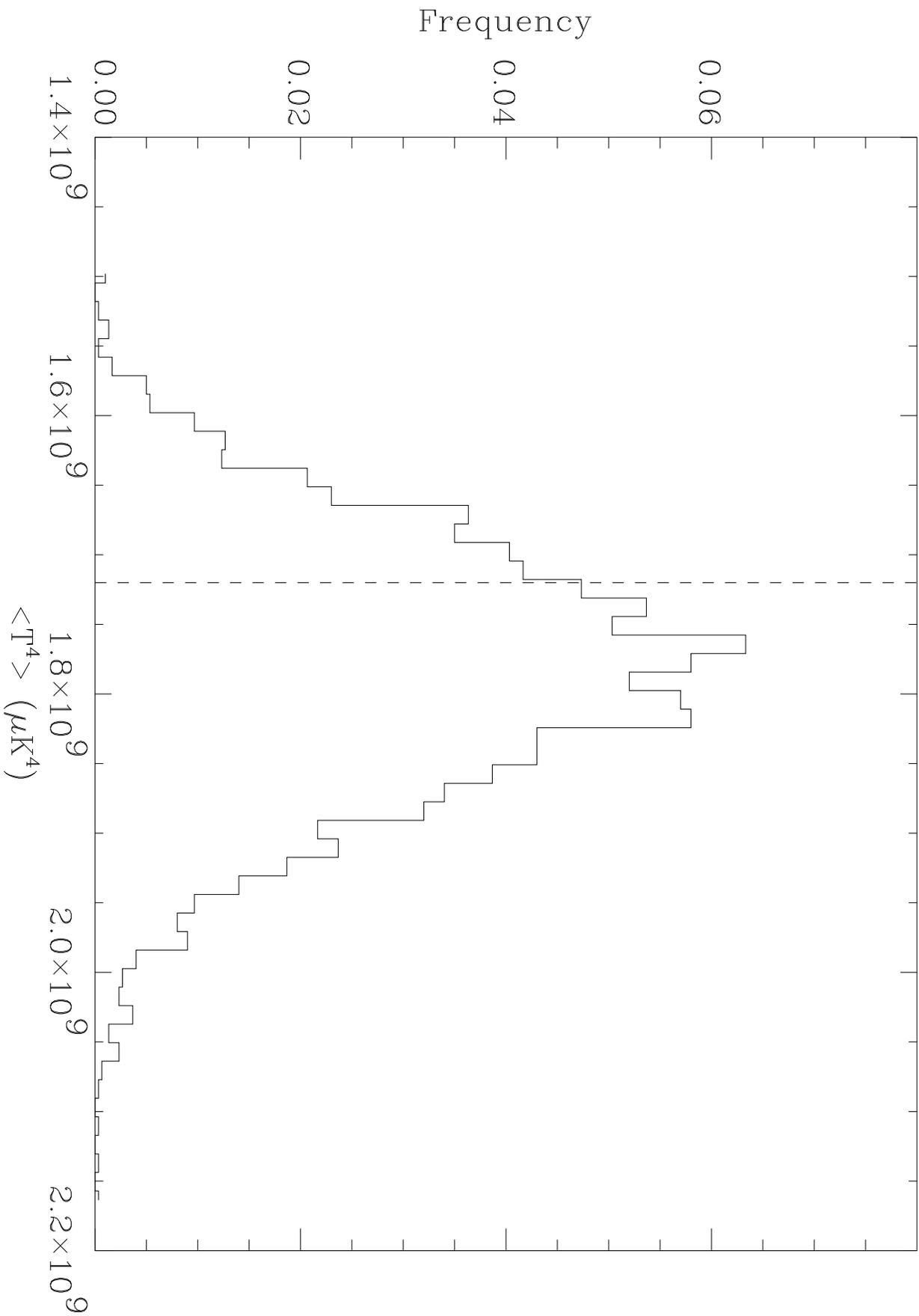

Figure 2d

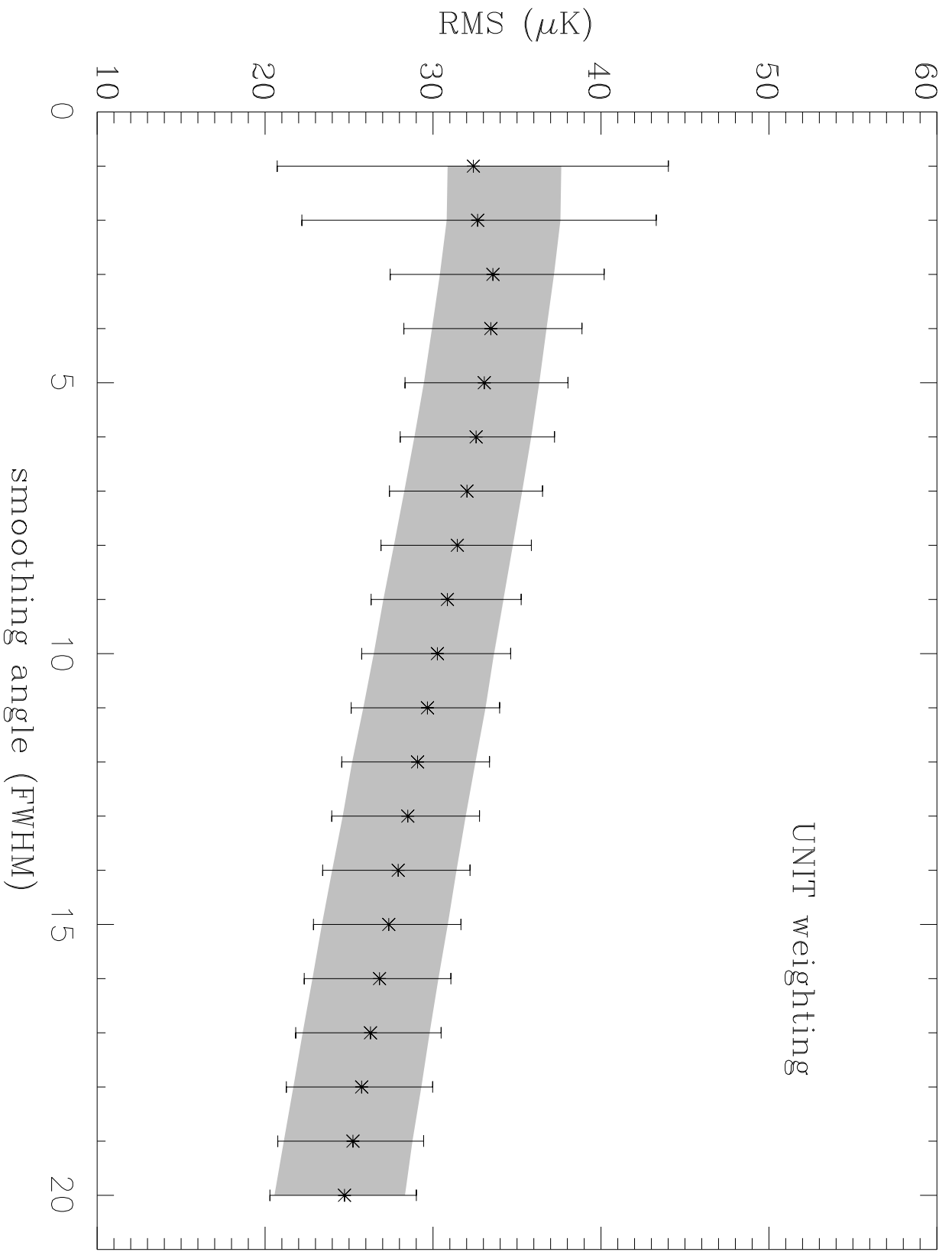

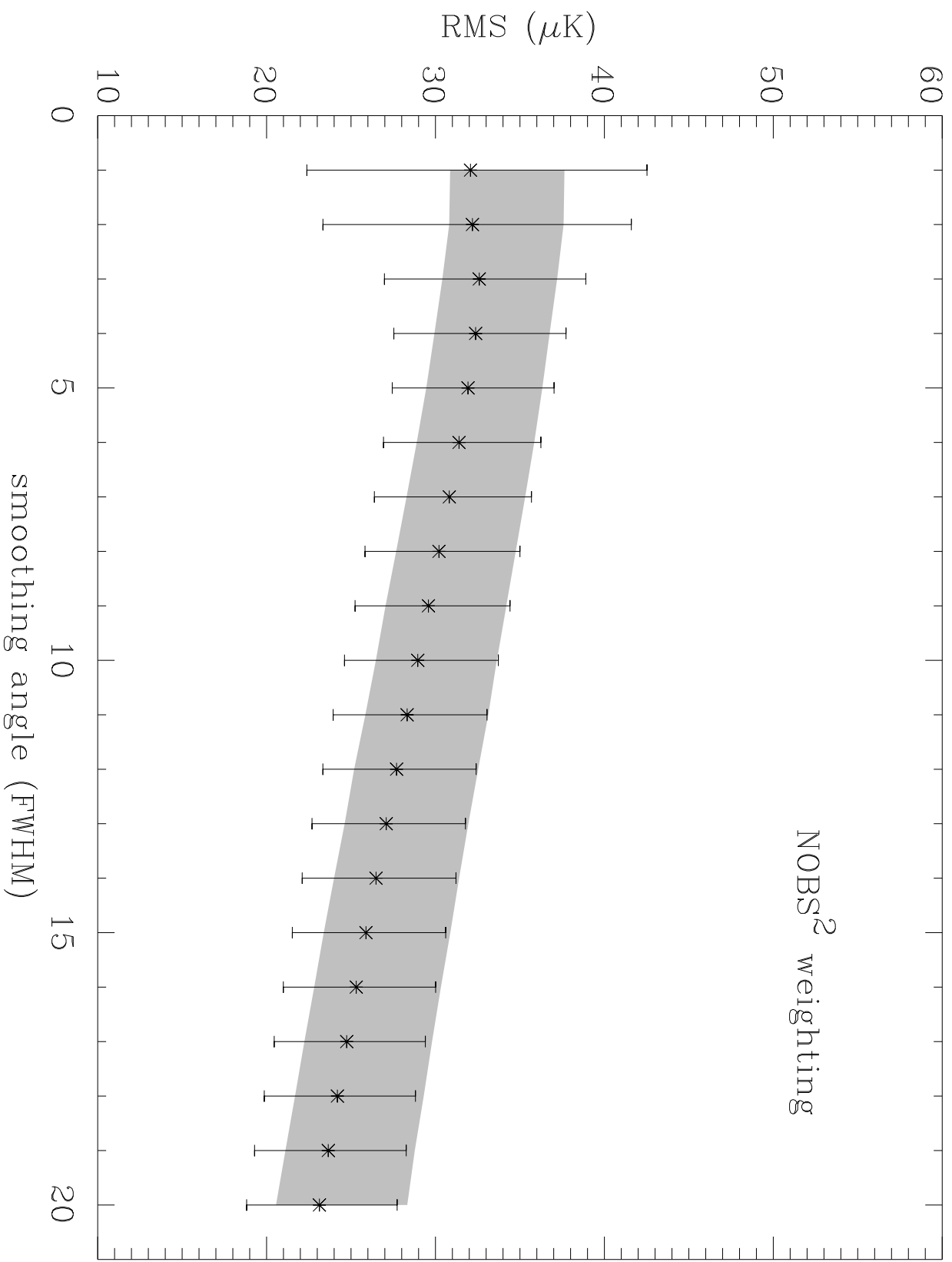

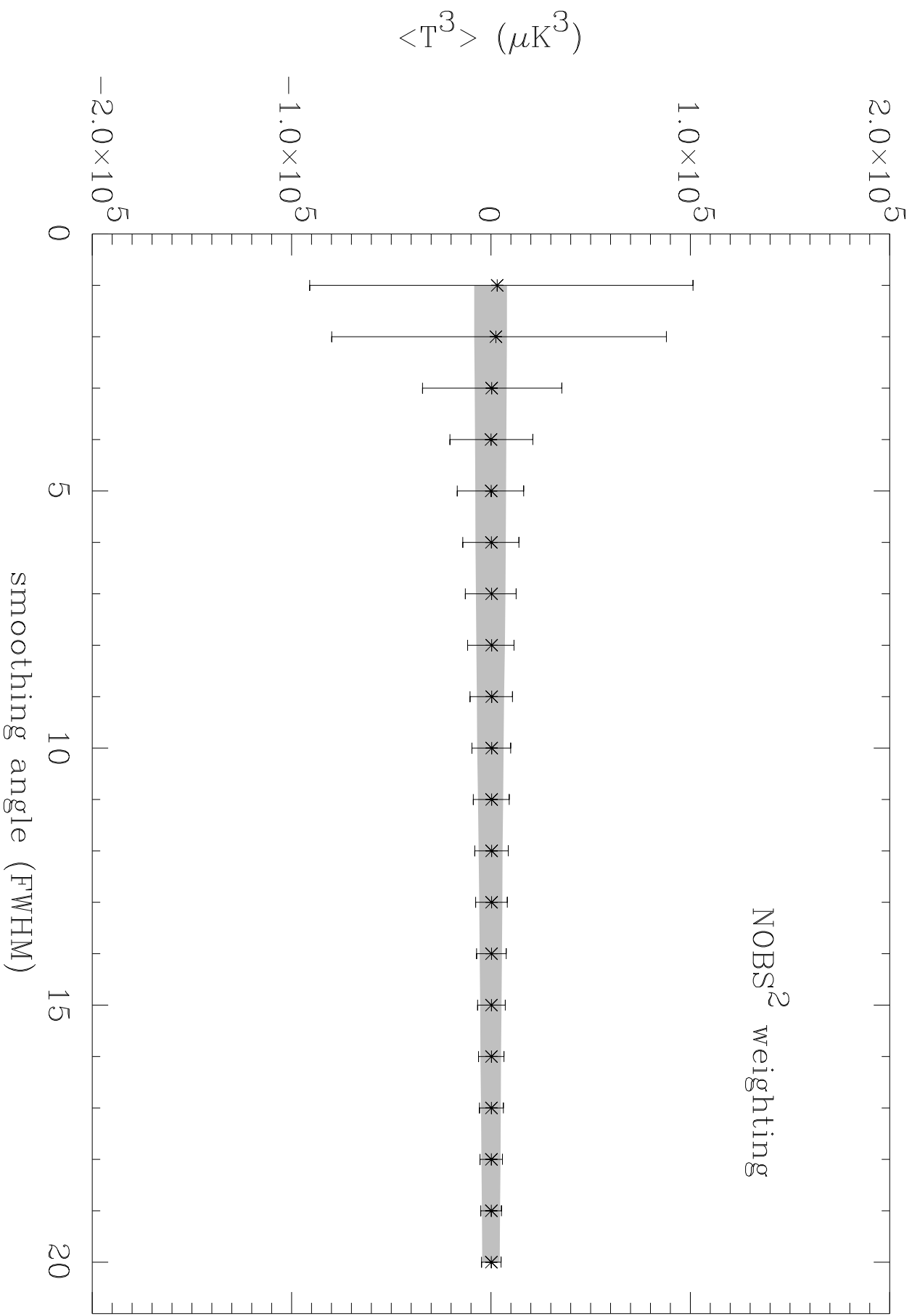

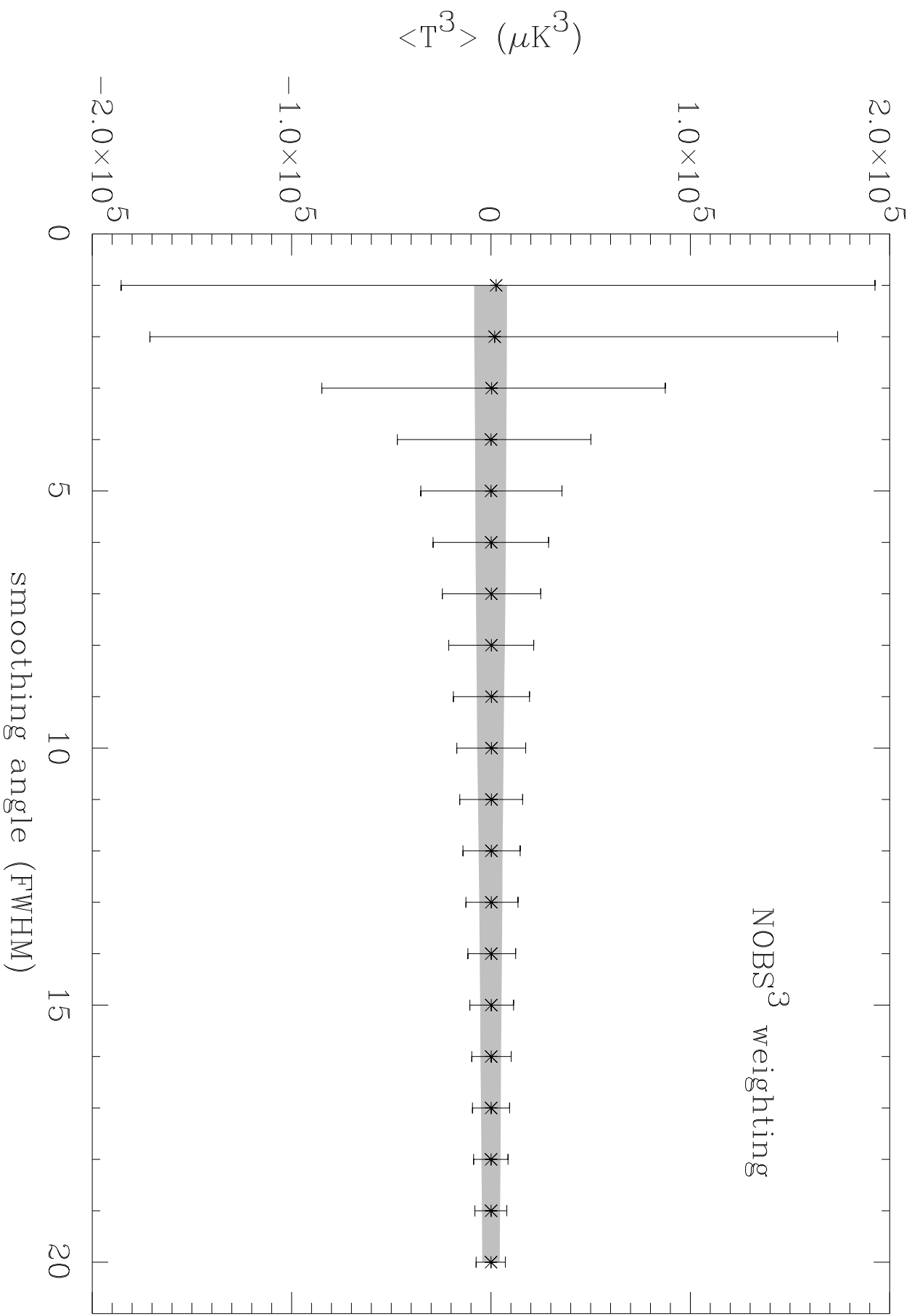

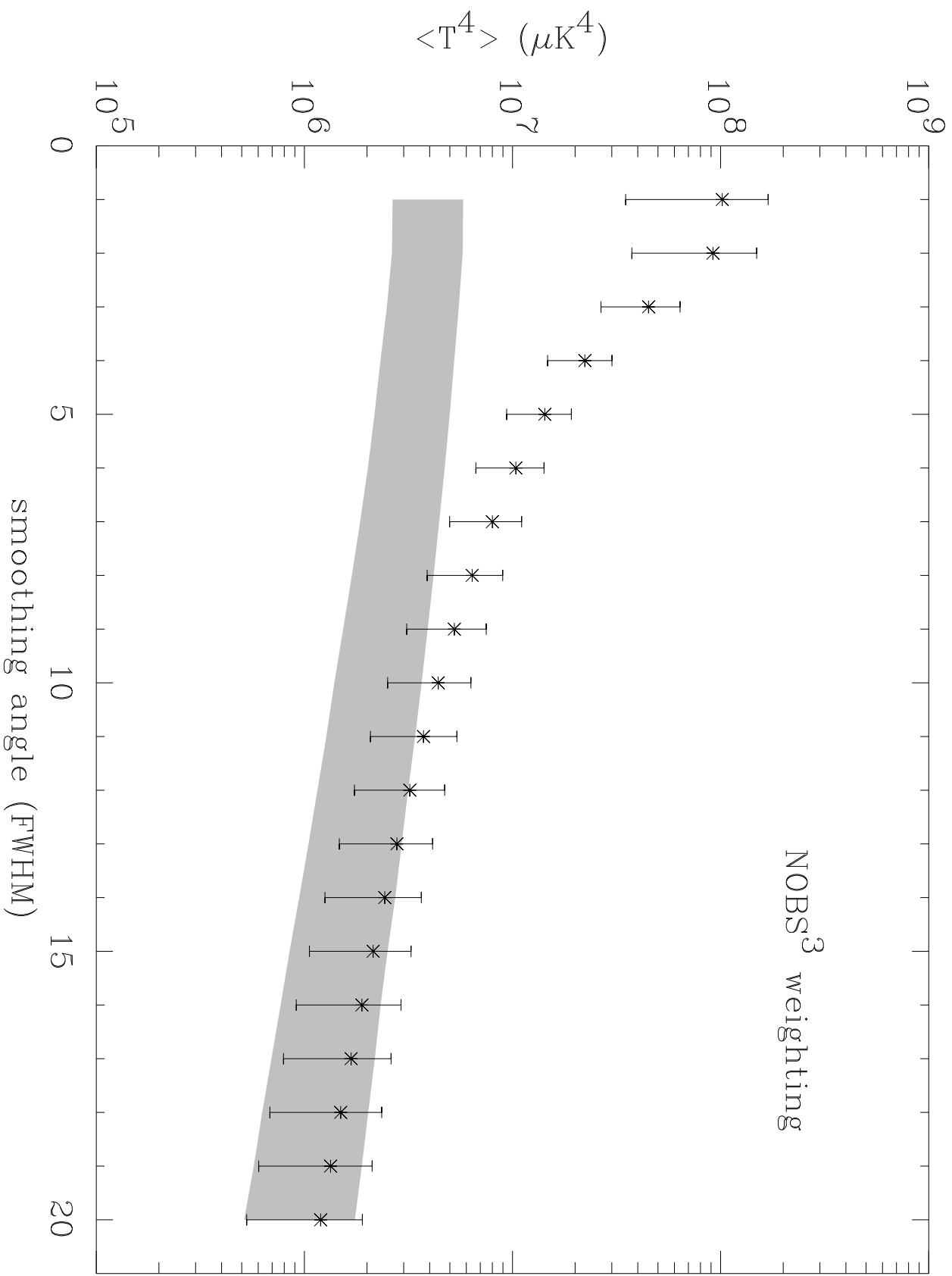

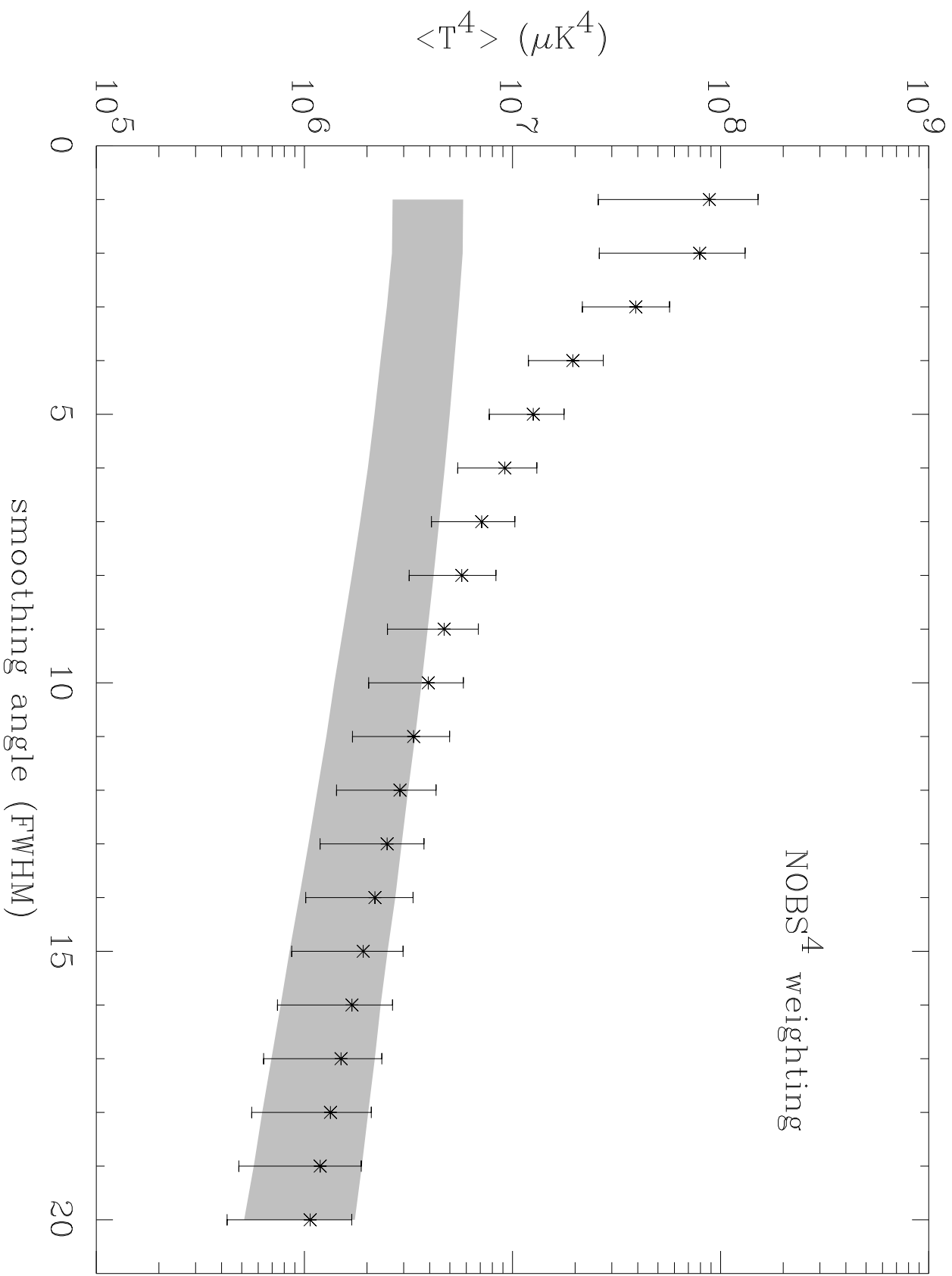

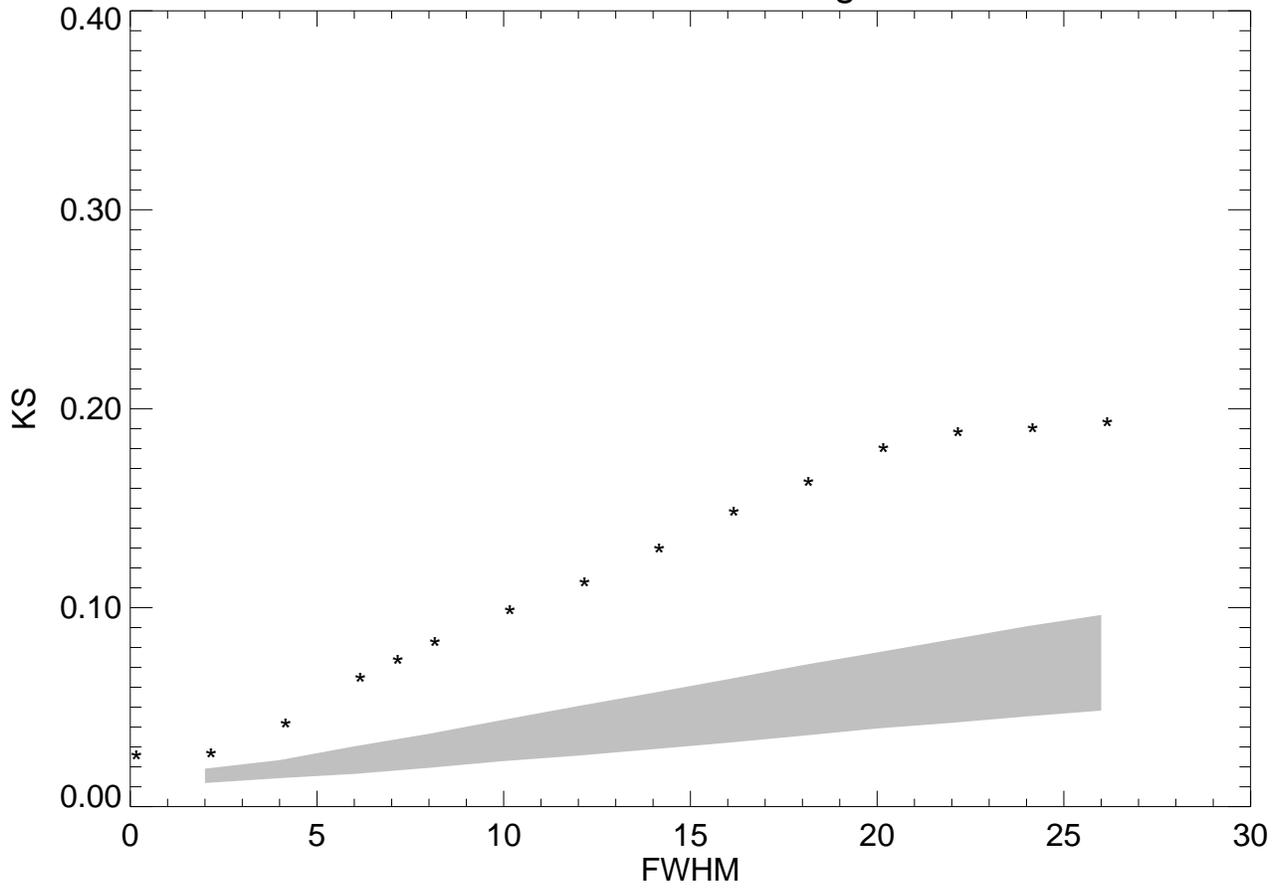

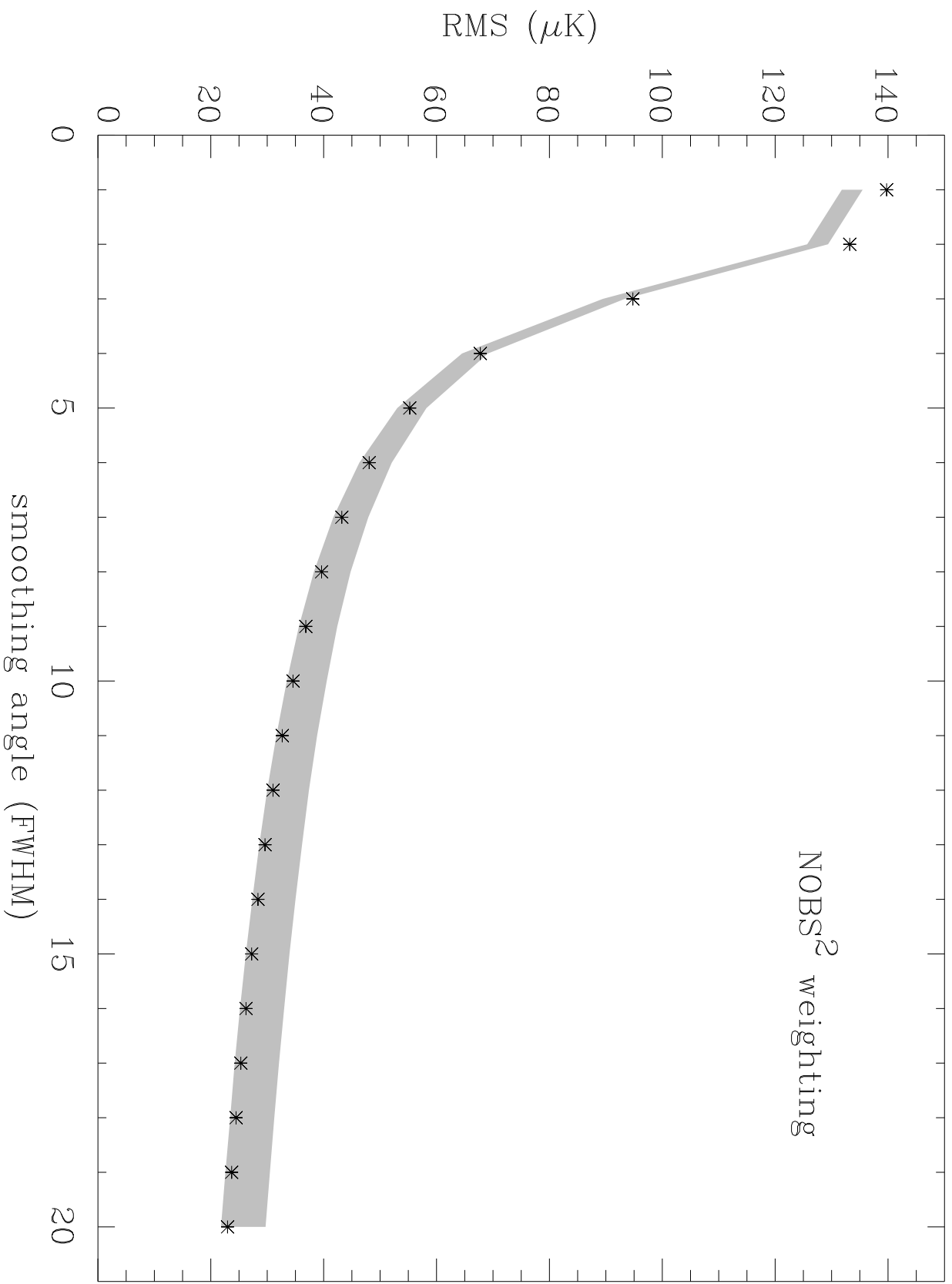

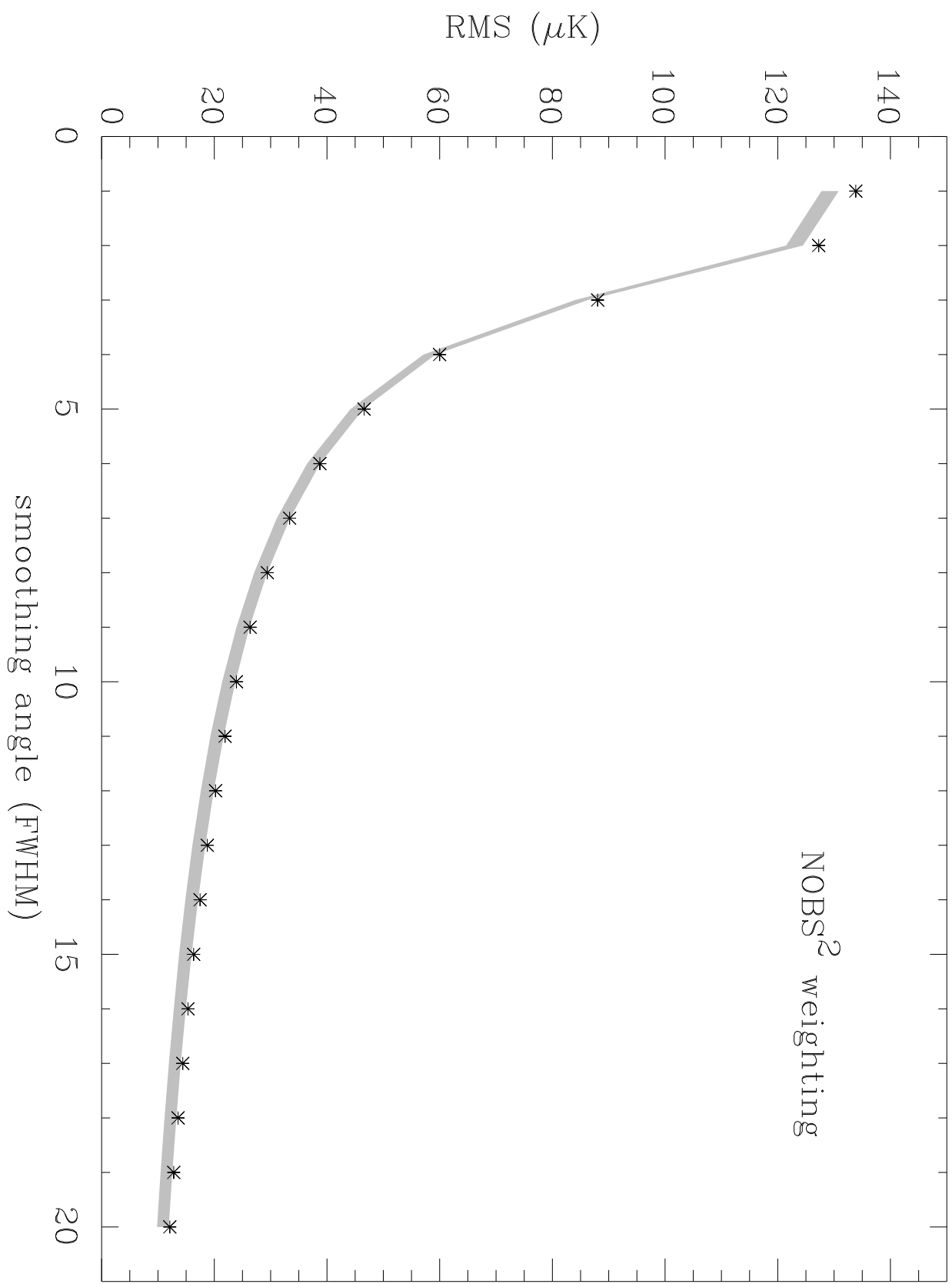

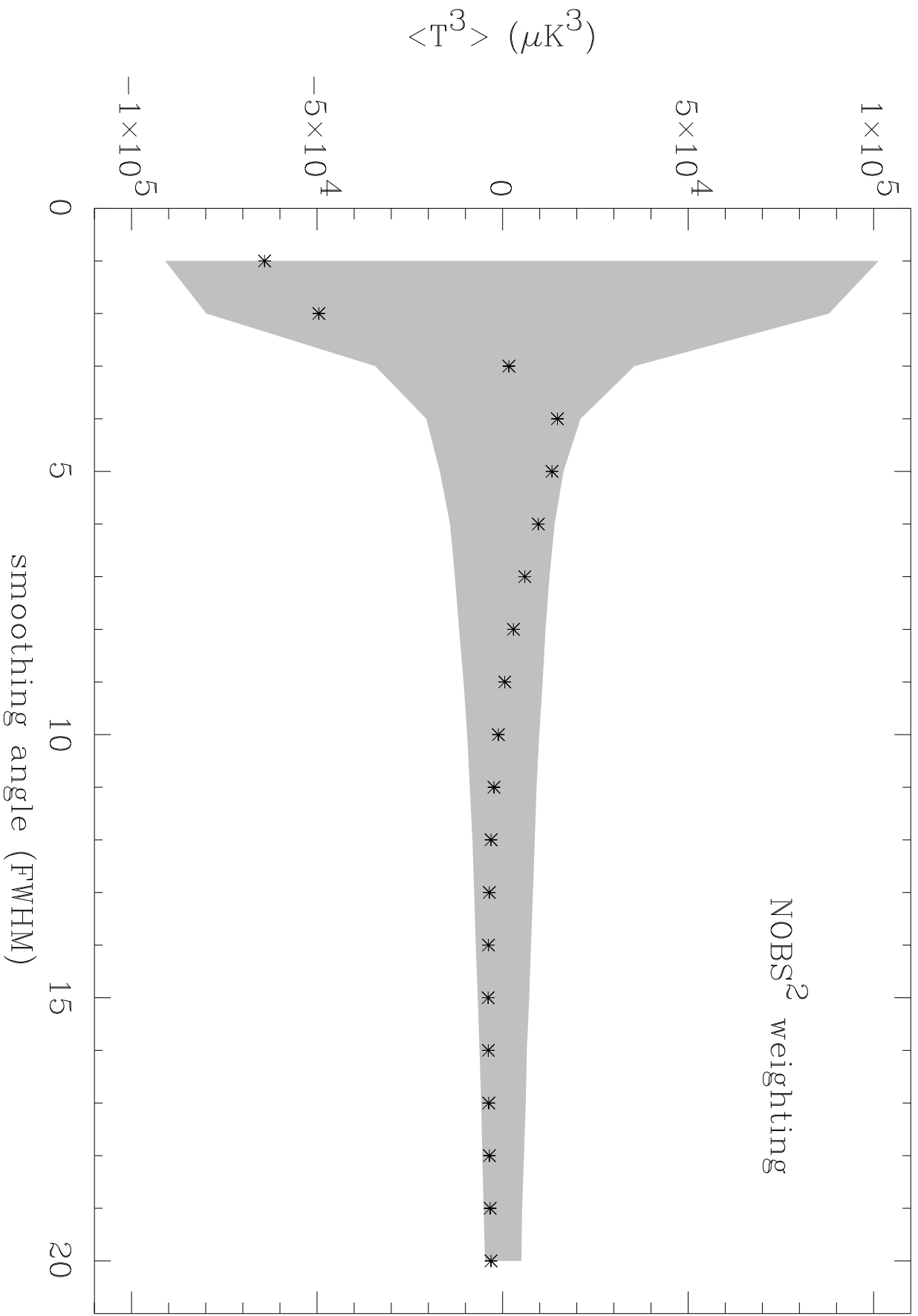

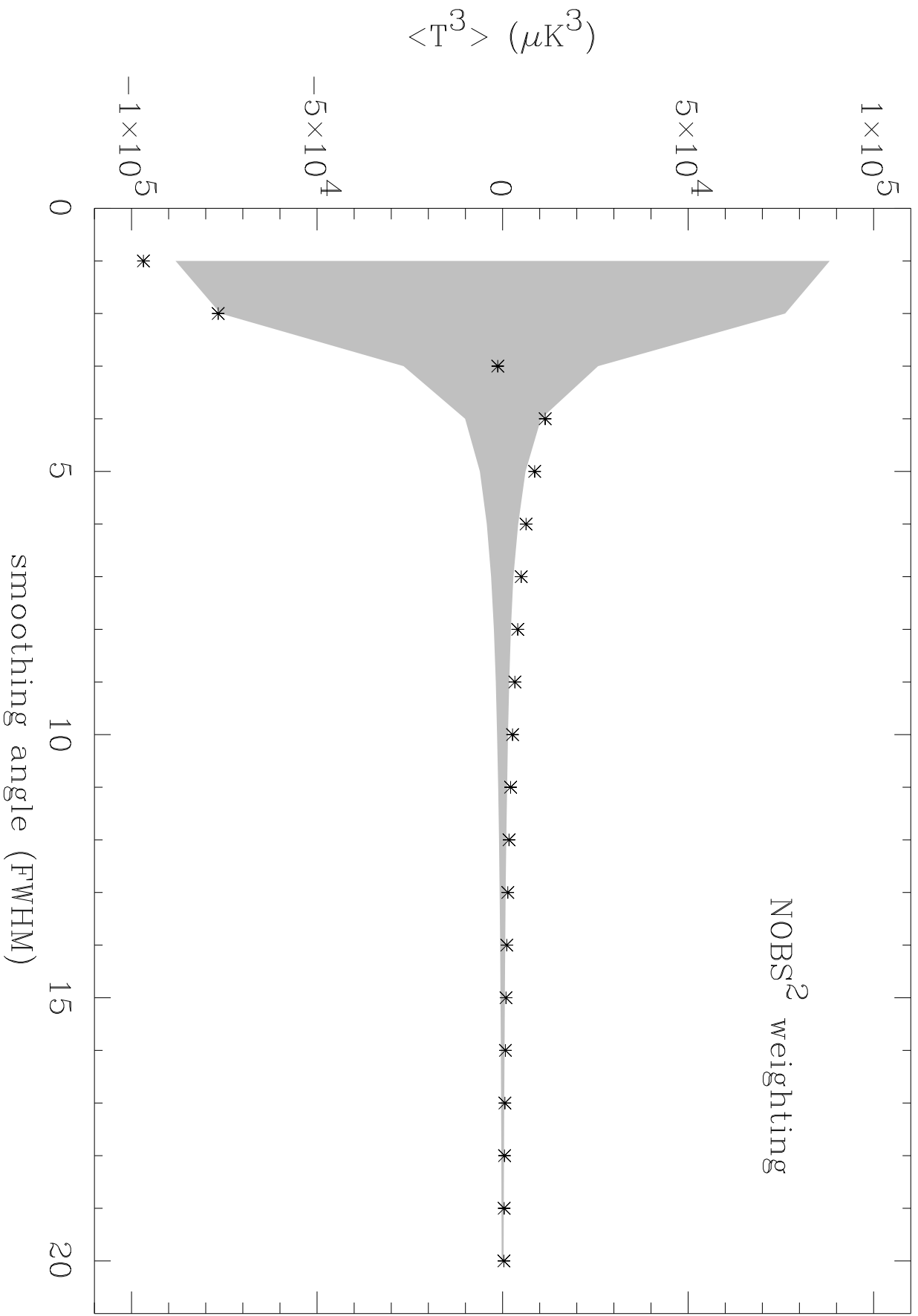

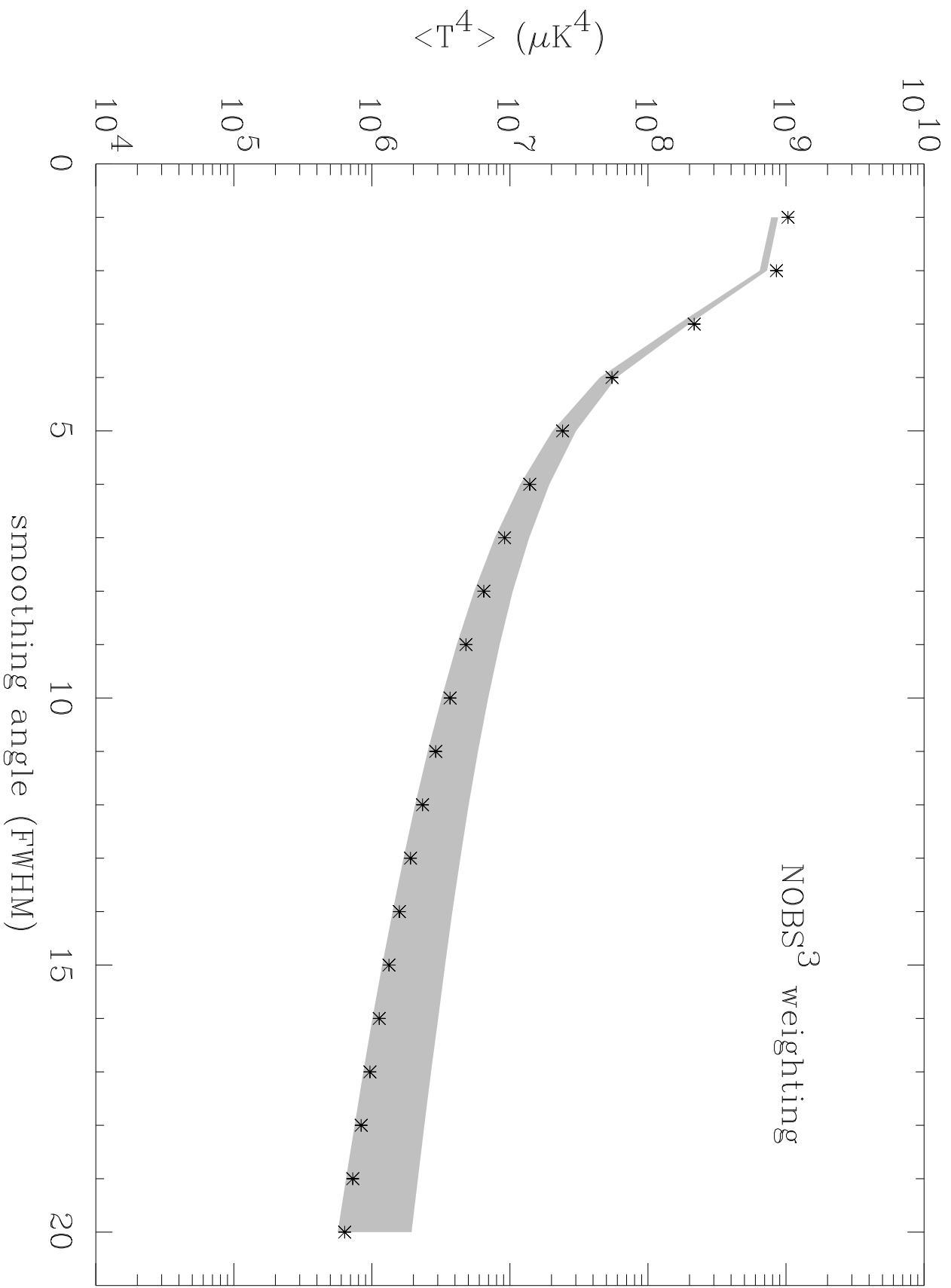

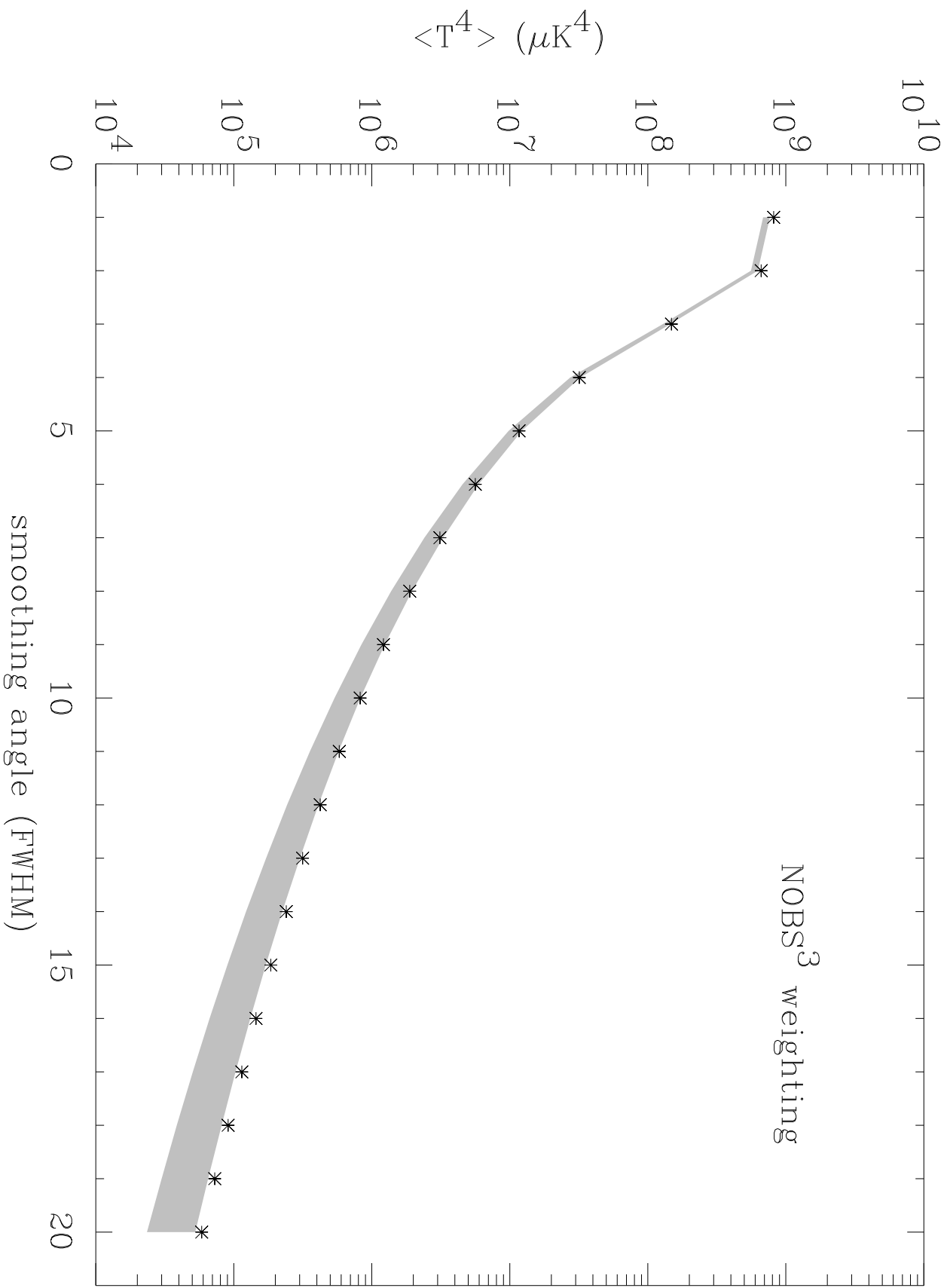

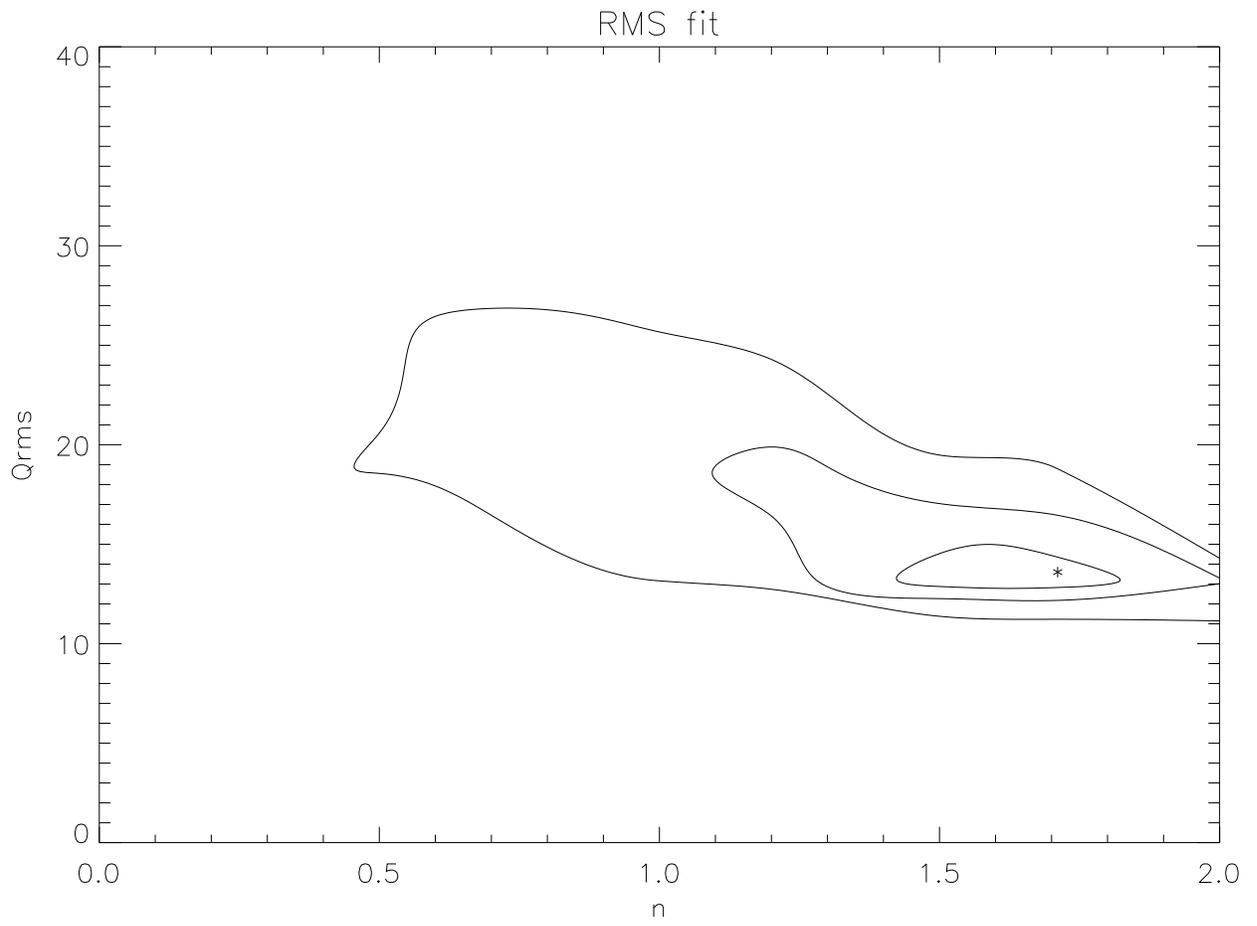

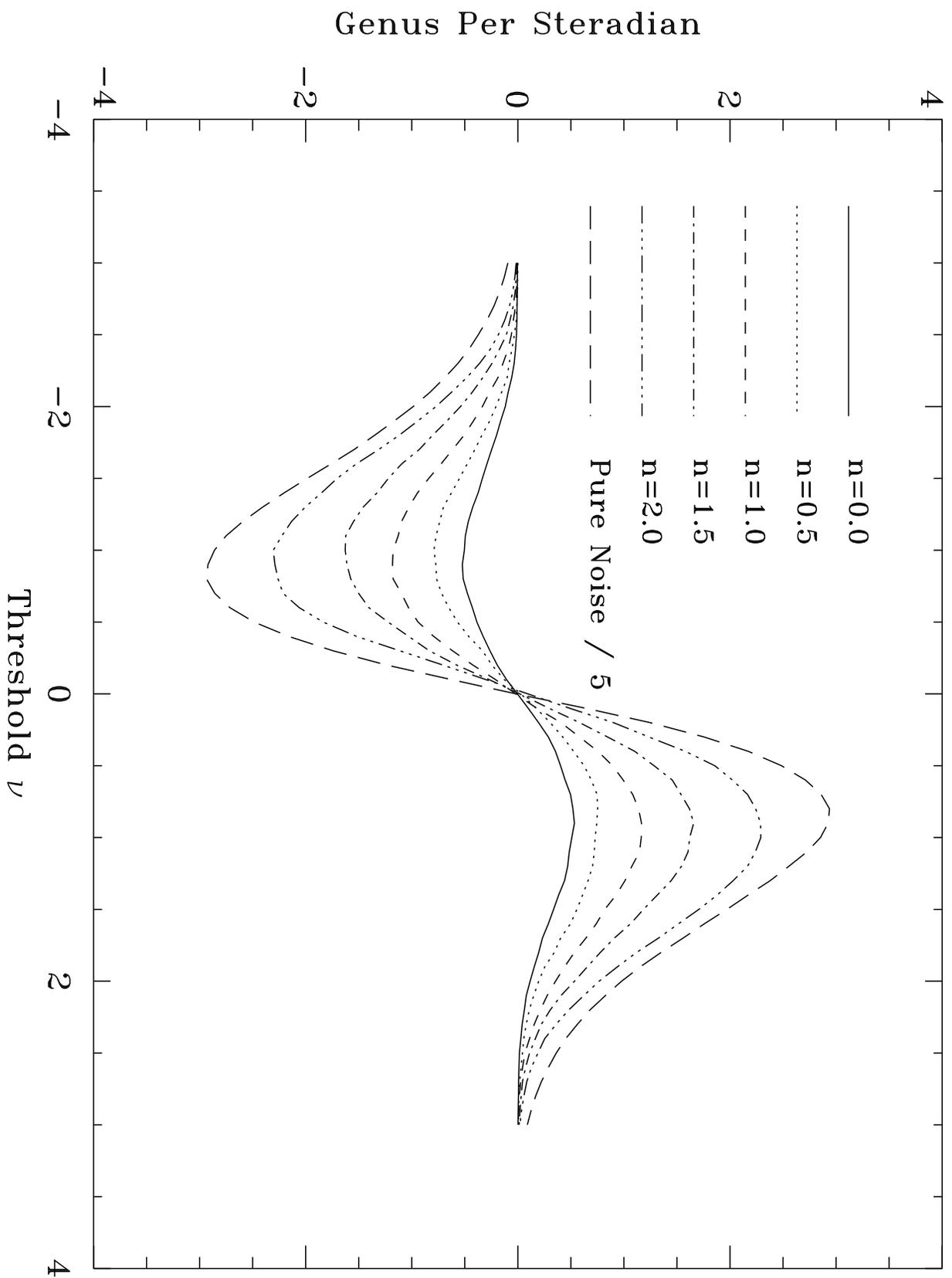

Figure 7

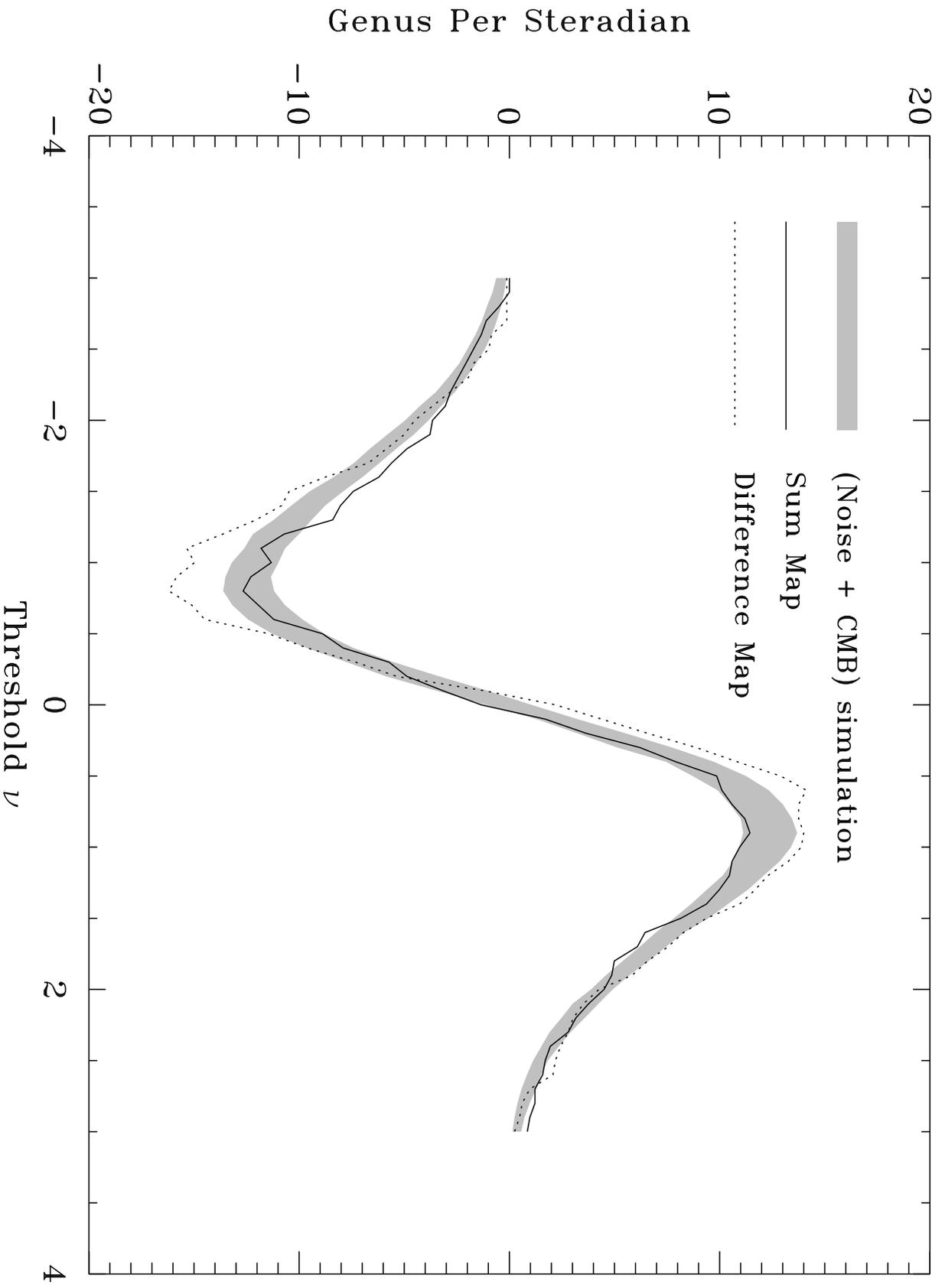

Figure 8a

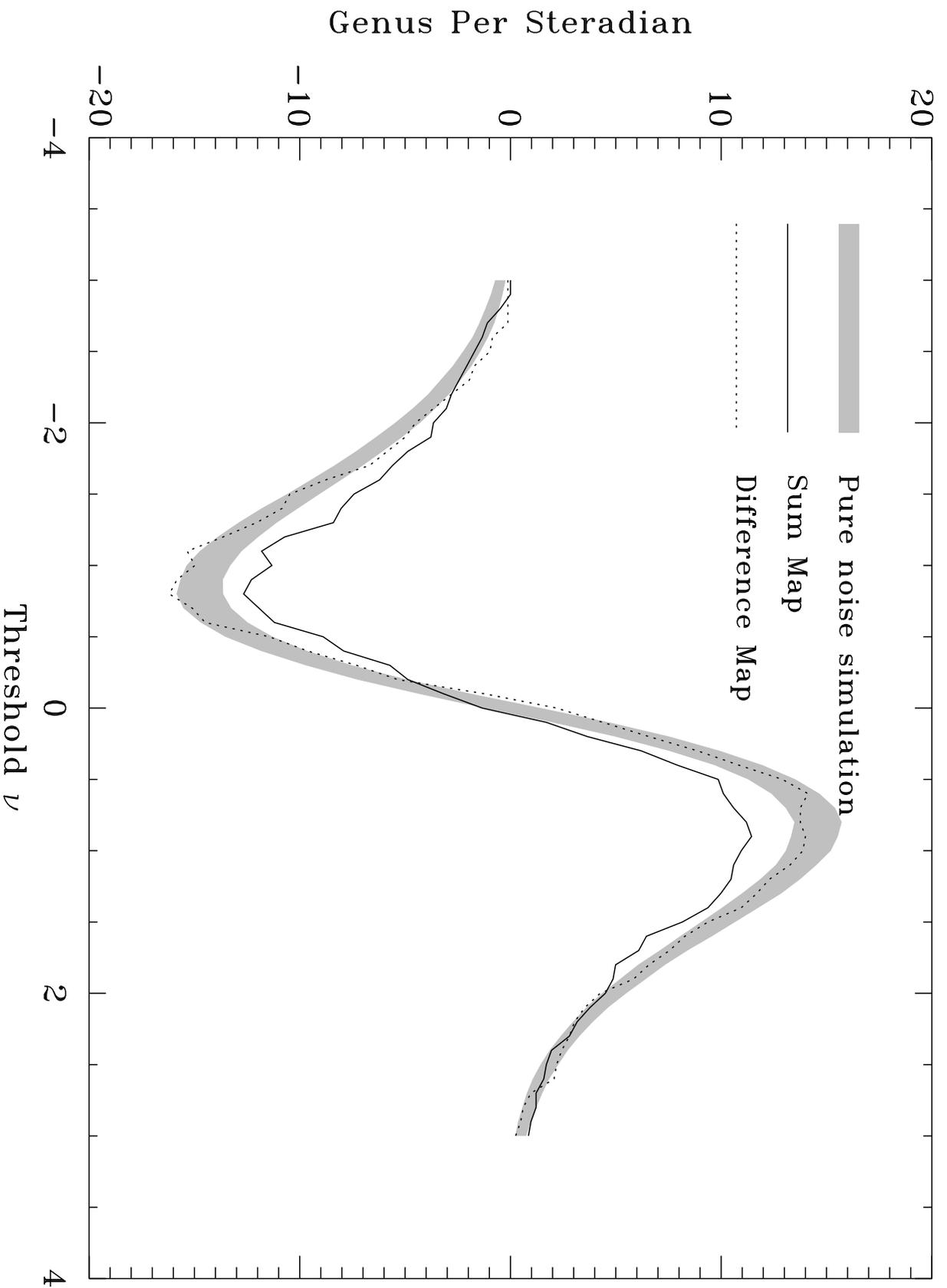

Figure 8b

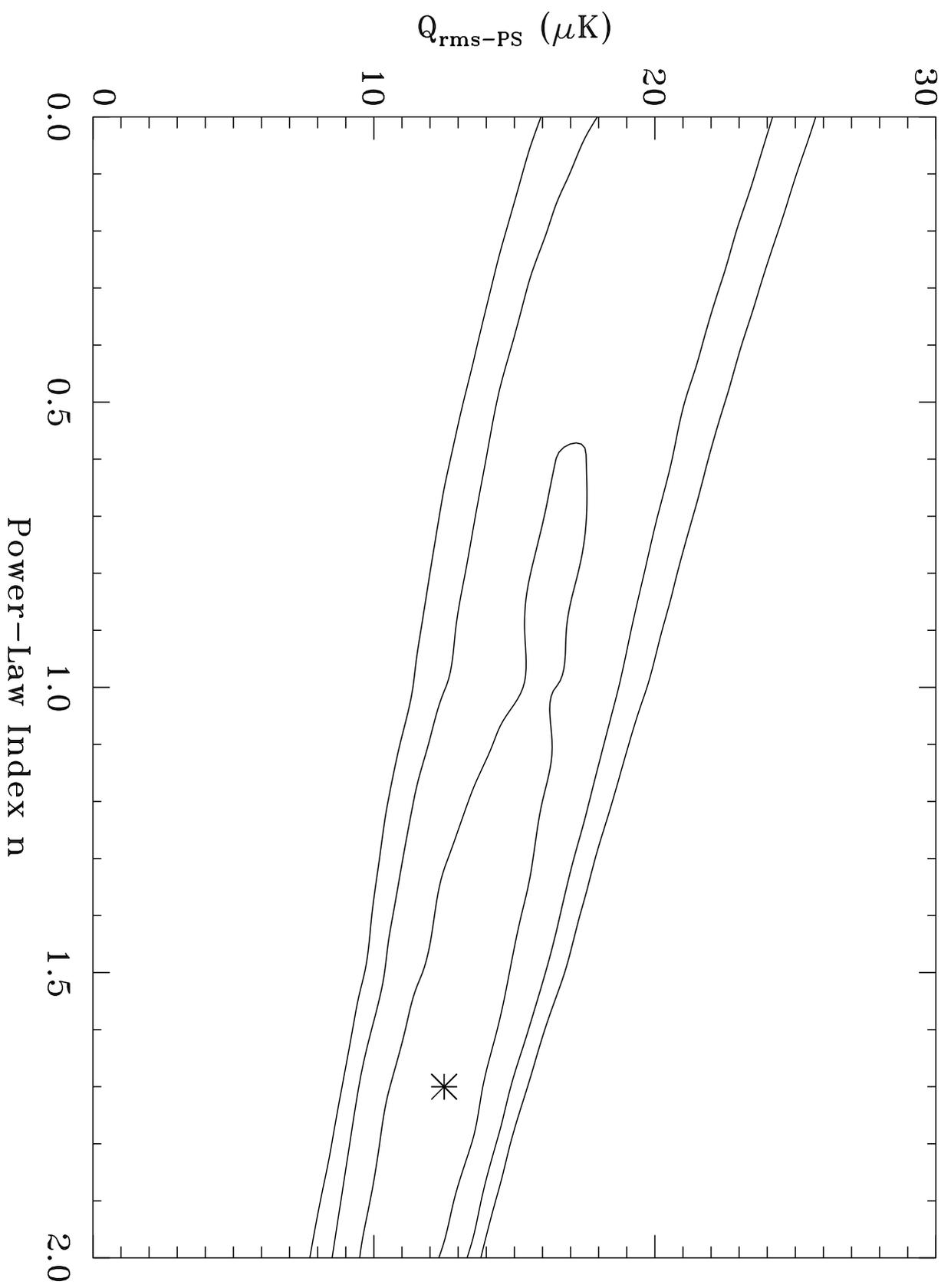

Figure 9

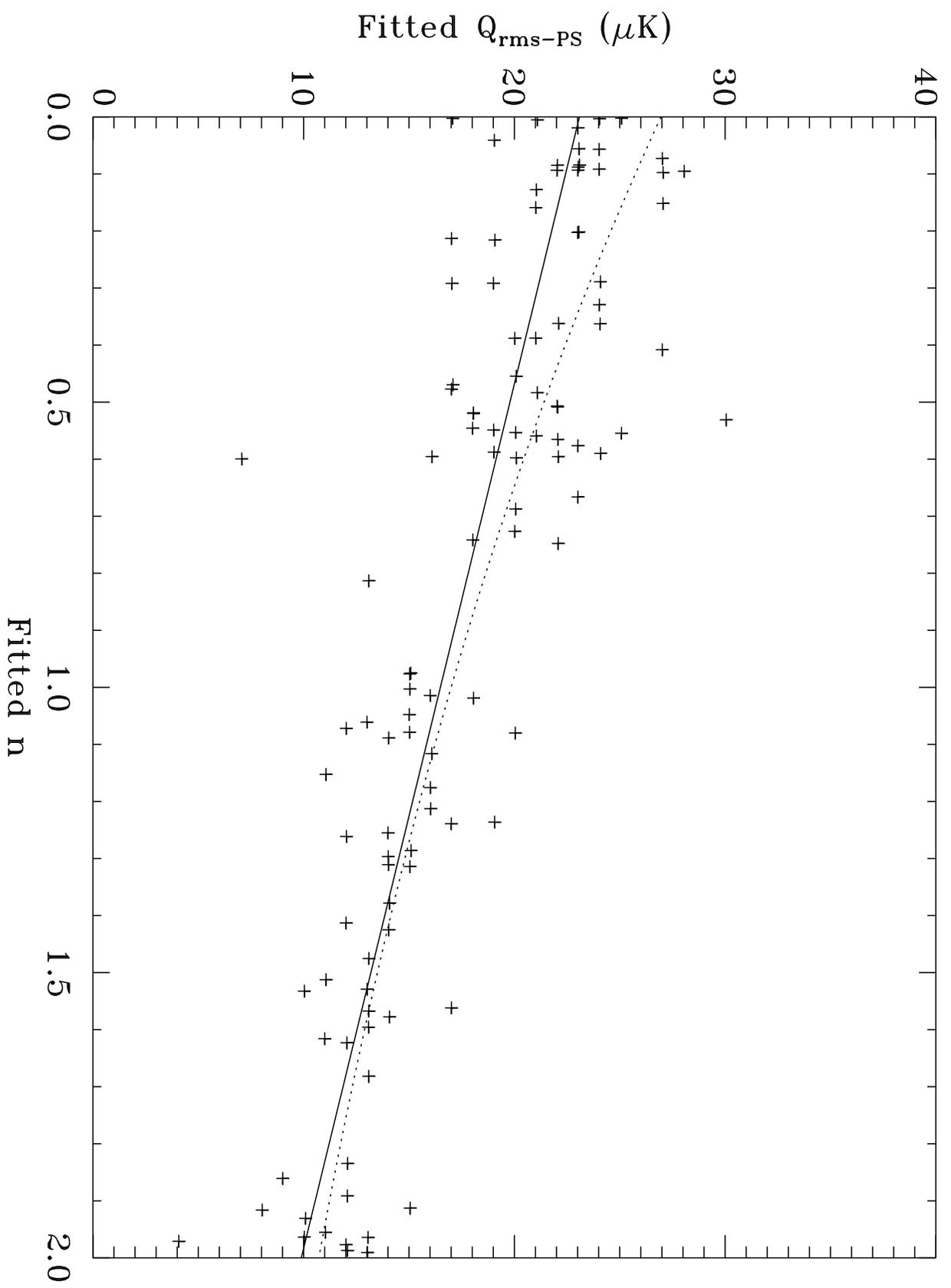

Figure 10